\documentclass[prd,12pt]{revtex4}

\usepackage{amsmath,amssymb} 
\usepackage[dvips]{graphicx} 
\usepackage[cp1251]{inputenc}

\newcommand\nn{\nonumber}
\newcommand\ba{\begin{eqnarray}}
\newcommand\ea{\end{eqnarray}}
\newcommand{\br}[1]{\left( #1 \right)}
\newcommand{\brs}[1]{\left[ #1 \right]}
\newcommand{\brf}[1]{\left\{ #1 \right\}}
\newcommand{\brm}[1]{\left| #1 \right|}
\newcommand{\Li}[1]{\mbox{Li}_{#1}}
\newcommand{\GeV}{~\mbox{GeV}}
\newcommand{\MeV}{~\mbox{MeV}}

\begin{document}

\begin{flushright}
DESY 09-115\\
November 2009
\end{flushright}

\title{Radiatively Corrected Lepton Energy Distributions in
Top Quark Decays $t \to b W^+ \to b(\ell^+ \nu_\ell)$
and $t \to b H^+ \to b(\tau^+ \nu_\tau)$
and single charged prong energy distributions from subsequent $\tau^+$ decays}

\author{Ahmed~Ali}
\email{ahmed.ali@desy.de}
\affiliation{Deutsches Elektronen-Synchrotron DESY, D-22607 Hamburg, Germany}

\author{Eduard~A.~Kuraev}
\email{kuraev@theor.jinr.ru}
\affiliation{JINR-BLTP, 141980 Dubna, Moscow region, Russian Federation}

\author{Yury.~M.~Bystritskiy}
\email{bystr@theor.jinr.ru}
\affiliation{JINR-BLTP, 141980 Dubna, Moscow region, Russian Federation}

\begin{abstract}
  We calculate the QED and QCD radiative corrections to the charged lepton energy
 distributions in the dominant semileptonic decays of the top quark
 $t \to b W^+ \to b(\ell^+ \nu_\ell)$ $(\ell=e, \mu, \tau) $ in the standard model
 (SM), and for the decay $t \to b H^+ \to b(\tau^+ \nu_\tau) $ in an extension of the SM
 having a charged Higgs boson $H^\pm$ with $m_{H^\pm} < m_t -m_b$.
 The  QCD corrections are calculated in the leading and next-to-leading
 logarithmic approximations, but the QED corrections are considered
 in the leading logarithmic approximation only. These corrections are numerically important
 for precisely testing the universality
of the charged current weak interactions in $t$-quark decays. As the $\tau^+$ leptons arising
from the decays $W^+ \to \tau^+ \nu_\tau$ and $H^+\to \tau^+ \nu_\tau$ are predominantly
left- and right-polarised, respectively, influencing the energy distributions
of the decay products in the subsequent decays of the $\tau^+$,
 we work out the effect of the radiative corrections on such distributions in
the dominant (one-charged prong) decay channels
 $ \tau^+ \to \pi^+ \bar{\nu}_\tau, \rho^+ \bar{\nu}_\tau, a_1^+ \bar{\nu}_\tau$
and $\ell^+ \nu_\ell \bar{\nu}_\tau$. The inclusive $\pi^+$ energy spectra in the
decay chains $t \to b(W^+,H^+) \to b (\tau^+ \nu_\tau) \to b (\pi^+ \bar{\nu}_\tau \nu_\tau +X)$
are calculated, which can help in searching for the induced $H^\pm$ effects at
the Tevatron and the LHC.
\end{abstract}

\vspace*{-5mm}
\maketitle

\section{Introduction}
\label{Introduction}

Top quark is now firmly established by the experiments CDF and D0 at the
$p\bar{p}$ collider Tevatron at Fermilab, with $m_t= 173.1 \pm 1.4$ GeV, decaying
dominantly through the mode $t \to b W^+ \to b (\ell^+\nu_\ell, q \bar{q}^\prime)$
\cite{:2009ec}.
At the Large Hadron Collider (LHC), expected to be operational shortly, one expects
a cross section $\sigma (p\,p \to t\bar{t}X) \simeq 1({\rm nb})$ for the
LHC centre of mass energy of 14 TeV~\cite{Langenfeld:2009tc}. With the nominal LHC
luminosity of $10^{33} ({\rm cm})^{-2} ({\rm sec})^{-1}$, one expects a $t \bar{t}$
pair produced per second. The $t\bar{t}$ production cross section for the 10 TeV
run of the LHC is estimated as about $0.4$ nb~\cite{Langenfeld:2009tc}, still large
 enough to undertake dedicated top quark physics.
  Thus, LHC is potentially a {\it top factory}, which will
allow to carry out precision tests of the SM and enhance the sensitivity of
beyond-the-SM effects in the top quark sector. Anticipating this,
a lot of theoretical work has gone into firming up the cross
sections for the $t\bar{t}$-pair and the single-top production at the
Tevatron and the LHC, undertaken in the form of higher order QCD corrections
\cite{Moch:2008ai,Moch:2008qy,Kidonakis:2008mu,Cacciari:2008zb}.
 Improved theoretical
calculations of the top quark decay width and distributions started a long time ago.
The leading order perturbative QCD corrections to the lepton energy spectrum in the
decays $t \to b W^+ \to b (\ell^+\nu_\ell)$ were calculated some thirty years
ago~\cite{Ali:1979is}.
Subsequent theoretical work leading to analytic derivations implementing the
 ${\cal O}(\alpha_s)$ corrections were published in~\cite{Corbo:1982ah,Altarelli:1982kh} and
corrected in  \cite{Jezabek:1988ja}. The order $\alpha_s$ contribution to the top quark decay
width dominates the radiative corrections (typically -8.5\%). The  $O(\alpha)$ electroweak
corrections contribute typically +1.55\% ~\cite{Denner:1990ns,Eilam:1991iz}, the finite
 $W$-width effect (-1.56\%) almost cancels the electroweak correction~\cite{Jezabek:1993wk}.
The next-to-leading order (NLO) QCD corrections in $\alpha_s$ (i.e., $\alpha_s^2$)
were computed as an expansion in
 $(M_W/m_t)^2$ in~\cite{Czarnecki:1998qc,Chetyrkin:1999ju}. These results were
 confirmed later by an independent analytic calculation
 in ~\cite{Blokland:2004ye,Blokland:2005vq},
and contribute about -2.25\% to the top quark decay width.

 Our main concern in this paper
are the lepton energy distributions from the decays $t \to b W^+ \to b (\ell^+\nu_\ell)$
(for $\ell^+ =e^+,\mu^+,\tau^+$), which are modified from their respective Born-level
distributions in a way specific for each charged lepton due to the QED corrections.
These (QED and QCD) radiative effects have to be taken into account to test the universality
of charged current weak interactions in the top quark sector. Another process which
breaks the charged lepton universality in the decays $ t \to b \ell^+ \nu_\ell$
is induced by charged Higgses $H^\pm$ (for $ m_{H^+} < m_t-m_b$) in the intermediate state,
$ t \to b H^+ \to b (\ell^+ \nu_\ell)$, which is expected to influence mainly the
final state $b \tau^+ \nu_\tau$ due to the $H^+\ell^+\nu_\ell$ couplings.
The leading order in $\alpha_s$ corrections to the polarized top quark decay into
$H^+b$ have been calculated in~\cite{Korner:2002fx}.
 We study the
effects of the radiative corrections on the $\tau^+$-energy distribution in the decay
$t \to b H^+ \to b \tau^+\nu_\tau$.

 Radiative (QED and QCD) corrections in the top
quark decays, such as $t \to b W^+ \to b \ell^+ \nu_\ell$, with $\ell^+=e^+, \mu^+, \tau^+$,
involve large logarithms due to the large fermion mass ratios. For example,
in the leading logarithmic approximation (LLA),
 one encounters the logarithmic terms
\ba
 L_e &=& \ln\br{\frac{m_t^2}{m_e^2}} \approx 25.4,
 \qquad
 L_\mu = \ln\br{\frac{m_t^2}{m_\mu^2}} \approx 14.8,
 \label{LDef}\\
 L_\tau &=& \ln\br{\frac{m_t^2}{m_\tau^2}} \approx 9.1,
 \qquad
 L_b = \ln\br{\frac{m_t^2}{m_b^2}} \approx 7.4, \nn
\ea
in the partial decay widths. Hence, in the LLA,  radiative
corrections to the partial widths lead to typically large effects
\ba
    \frac{\alpha}{\pi} L_e \approx 6.2~\%,
    \qquad
    \frac{\alpha}{\pi} L_\mu \approx 3.6~\%,
    \qquad
    \frac{\alpha}{\pi} L_\tau \approx 2.1~\%,
    \qquad
    \frac{\alpha_s}{\pi} L_b \approx 23~\%.
\ea
They are included together with the non-logarithmic terms in the estimates undertaken in
the fixed order (in $\alpha$ or $\alpha_s$) calculations. However, to get perturbatively
reliable results, all terms of the type $(\frac{\alpha}{\pi})^n\ln\br{\frac{m_t^2}{m_e^2}}^n$
in the decay $t \to b e^+ \nu_e$, for example,
have to be summed up (the re-summed leading log approximation LLA), as well as
$(\frac{\alpha}{\pi})^n\ln\br{\frac{m_t^2}{m_e^2}}^{n-1}$ (the next-to-leading log
approximation NLLA).  Using the well-studied case of the QED radiative corrections
to the purely leptonic decays $\mu^- \to \nu_\mu e^- \bar{\nu}_e$,
we show that the structure function (SF) approach~\cite{Lipatov:1974qm,Altarelli:1977zs}
(based on the factorisation hypotheses \cite{Collins:1984kg}) is the appropriate framework
to resum such terms,
enabling us to derive the electron energy spectrum with the
radiative corrections taken into account to all orders of the large logarithms.
 As a warm-up exercise, and also to set our notations,  we reproduce the well-known results for the QED corrections to the muon decay
$\mu ^- \to e^- \bar{\nu}_e \nu_\mu$
\cite{Berman:1958ti,Kinoshita:1957zz,Arbuzov:2002pp,Arbuzov:2002cn}
and generalise it to all orders of perturbation theory by summing up the leading logs
 $(\frac{\alpha}{\pi}
\ln (m_\mu^2/m_e^2 ))^n$ (see Section~\ref{MuonSection}). In this context, we also discuss the
 polarised muon decay case. The SF approach is applied next to the semileptonic
decays of the top quark $t \to bW^+ \to b (\ell^+ \nu_\ell)$, where the QCD and QED radiative
corrections to the Dalitz (double differential) and inclusive lepton energy distributions are worked
out. In this, the QCD-corrected energy distributions are derived in the re-summed leading logarithmic
and next-to-leading logarithmic approximations, but the QED corrections to these distributions are
calculated in the leading logarithmic approximation only.
This is discussed in detail in Section~\ref{TopSection}.

In many extension of the SM, the Higgs sector of the SM is enlarged, typically by adding an
extra doublet of complex Higgs fields. After spontaneous symmetry breaking, the two
scalar Higgs doublets $\Phi_1$ and $\Phi_2$  yield three physical neutral Higgs bosons
($h,H,A $) and a pair of charged Higgs bosons ($H^\pm$).
 If $m_{H^+} \leq m_t -m_b$,   one expects measurable effects in the top quark decay width and decay
distributions
due to the $H^\pm$-propagator contributions, which are potentially large in the decay
chain $t \to b H^+ \to b (\tau^+\nu_\tau)$. The two parameters which determine the
branching ratio for this decay are $m_{H^\pm}$ and the quantity called $\tan \beta$,
defined as $\tan \beta \equiv v_2/v_1$, where $v_1$ and $v_2$ is the vacuum expectation
value of $\Phi_1$ and $\Phi_2$, respectively.
Of particular interest is the parameter space with large
$\tan \beta$ (say, $\tan \beta > 20$) and $m_{H^\pm} \leq 150$ GeV. This mass range is
already excluded (for almost the entire $\tan \beta$ values of interest)  in the
so-called two-Higgs-doublet-models 2HDM due to the lower bound on $m_{H^\pm}$
of 295 (230) GeV at the 95\%(99\%) C.L. from the experimental measurements of the branching ratio
${\cal B}(B \to X_s \gamma)$~\cite{Amsler:2008zzb},  and the order $\alpha_s^2$ estimates of
 this quantity in the SM~\cite{Misiak:2006zs}. However, this bound applies only to the 2HDM of
type II, in which the Higgs doublets $\Phi_1$ and $\Phi_2$ couple only to the right-handed
down-type fermions $(d_{iR}, \ell_{iR})$ and the up-type fermions $(u_{iR},\nu_{iR})$,
 respectively. In the minimal supersymmetric standard model (MSSM), one has a type II 2HDM
sector in addition to the supersymmetric particles, in particular
the charginos, stops and  gluinos. Their contributions could, in
principle, cancel that of the charged Higgs bosons in the
 $B \to X_s + \gamma$ decay rate. Hence,
the 2HDM-specific constraint on $m_{H^\pm}$ from ${\cal B}(B \to X_s \gamma)$ is not applicable
 in the MSSM. In our opinion,
 the natural embedding of the extra Higgs doublet is in a supersymmetric theory,
 and hence we
will ignore the lower bound on $m_{H^\pm}$ from ${\cal B}(B \to X_s \gamma)$.
 A model-independent lower bound on $m_{H^\pm}$ exists from the non-observation of the
charged Higgs pair production at  LEPII, yielding
 $m_{H^\pm} > 79.3$ GeV at 95\% C.L.~\cite{Amsler:2008zzb}, which we shall use in our numerical
 analysis. Thus, a charged Higgs having a mass in the
range $80~{\rm GeV} \leq m_{H^\pm} \leq 160$ GeV is a logical possibility and its
effects should be searched for in the decays $t \to b H^+\to \tau^+ \nu_\tau$.
 A beginning along these lines
has already been made at the Tevatron~\cite{Abbott:1999eca,Abazov:2001md,Abulencia:2005jd},
 but a definitive search will be carried out only at
the LHC~\cite{:2008zzm,Bayatian:2006zz}.
 We work out the effects of the radiative corrections
to the lepton energy spectra in the decays $t \to bH^+ \to b (\tau^+ \nu_\tau)$ in
 Section~\ref{TopHiggsDecays}.

The $\tau^+$ leptons arising
from the decays $W^+ \to \tau^+ \nu_\tau$ and $H^+\to \tau^+ \nu_\tau$ are predominantly
left- and right-polarised, respectively. Polarisation of the $\tau^\pm$  influences the
 energy distributions
 in the subsequent decays of the $\tau^\pm$.
 Strategies to enhance the $H^\pm$-induced effects in the decay
 $t \to b W^+ \to b (\tau^+ \nu_\tau)$,  based on the
polarisation of the $\tau^+$ have been discussed at length in the existing literature
~\cite{Hagiwara:1989fn,Bullock:1991fd,Rouge:1990kv,Bullock:1992yt,Raychaudhuri:1995cc}.
 We work out the effect of the radiative corrections on such distributions in
the dominant (one-charged prong) decay channels
 $ \tau^+ \to \pi^+ \nu_\tau, \rho^+ \nu_\tau, a_1^+ \nu_\tau$
and $\ell^+ \bar{\nu}_\ell \nu_\tau $.  To implement this, we again use the
 SF approach~\cite{Kuraev:1985hb}.  In particular, the inclusive $\pi^+$
 energy spectrum  in the decay chain
 $t \to b (W^+,H^+) \to b (\tau^+ \nu_\tau) \to b (\pi^+ \bar{\nu}_\tau \nu_\tau +X)$,
and likewise for the decay chain of the $\bar{t}$ quark,
 can be used to search for the induced effects of the
$H^\pm$ at the LHC and Tevatron. Details are given in Section~\ref{TauSpectrumModification}
 and in Appendix A.

To get the relative normalisation of the decay width $t \to bH^+$ with respect to the SM
decay width $t \to bW^+$, one has to take into account the loop corrections (quantum soft SUSY-breaking
effects). These quantum effects on $t \to b H^+$ have been worked out in the context of
the minimal supersymmetric standard model MSSM in
a number of detailed studies (see, for example~\cite{Coarasa:1996qa,Carena:1999py}), and the
bulk of them can be implemented by
modifying the $b$-quark mass, $m_b^{\rm corrected}=m_b/(1+\Delta_b)$. The specific values of
$\Delta_b$ depend on the supersymmetric mass spectrum, and can be calculated using
FeynHiggs~\cite{Hahn:2008zzd}, given this spectrum. The influence of these corrections on the branching
 ratio for the
 decay $t \to b H^+$ have been recently updated in~\cite{Sopczak:2009sm}, predicting
 $BR(t \to b H^+) \geq 0.1$ for $m_{H^+} \leq 110$ GeV in the large-$\tan \beta$
 region ($\tan \beta > 40$). We shall pick a point in the $(\tan \beta - m_{H^+})$ plane from
this study,  allowed by all current searches, for the sake of illustration.
 We summarise our results in Section~\ref{TopHiggsSummary}.

\section{Muon decay: A warm-up Exercise}
\label{MuonSection}

We start by discussing the electron energy spectrum in
$\mu\to e\bar \nu_e \nu_\mu$ decay. In the Born
approximation, this spectrum is given by the following formula \cite{Okun:1982ap}:
\ba
    \frac{d\Gamma_B}{dx} = 6 \Gamma_\mu
    \brs{2x^2\br{1-x} - \frac{4}{9} \rho x^2 \br{3-4x}},
    \label{MuonSpectrumBorn}
\ea
where $x = 2E_e/m_\mu$ is the energy fraction of final electron,
$\rho$ is the well-known Michel parameter \cite{Michel:1949qe} and
$\Gamma$ is the total decay width:
\ba
    \Gamma_\mu
    =
    \frac{G_F^2 m_\mu^5}{192\pi^3}~,
\ea
where $G_F$ is the Fermi coupling constant.
Using the SF approach \cite{Lipatov:1974qm,Altarelli:1977zs},
we can derive the electron-energy spectrum with the
radiative corrections taken into account to all orders of the large logarithm:
\ba
    \frac{d\Gamma_{RC}}{dx}
    &=&
    \int\limits_{x}^1 \frac{dy}{y}
    D\br{\frac{x}{y},\beta}
    \frac{d\Gamma_B}{dy}
    \br{1+\frac{\alpha}{2\pi} K\br{y}},
    \label{MuonSpecrtumWithRC}\\
    &&
    \beta=\frac{\alpha}{2\pi}\br{L-1},
    \qquad
    L=\ln\br{\frac{m_\mu^2}{m_e^2}}\approx 10, \nn
\ea
where $\frac{\Gamma_{B}}{dx}$ is the electron spectrum in the Born approximation
(\ref{MuonSpectrumBorn}) which is considered as the
hard sub-process. $D\br{x,\beta}$ is the so-called
structure function, which describes the virtual and real photon
emission in the leading logarithmic approximation
and has the form \cite{Kuraev:1985hb}:
\ba
    D\br{x,\beta} &=&
    \delta\br{1-x} + \beta P^{(1)}\br{x} +
    \frac{1}{2!} \beta^2 P^{(2)}\br{x} + \cdots~.
    \label{SF}
\ea
The quantities $P^{(n)}\br{x}$ are the  kernels
of the  evolution equations which
are defined by the following relations:
\ba
    P^{(1)}\br{x} &=&
    \br{\frac{1+x^2}{1-x}}_+
    =
    \lim_{\Delta\to 0}
    \brs{
        \frac{1+x^2}{1-x} \theta\br{1-x-\Delta}
        +
        \br{2\ln\br{\Delta} + \frac{3}{2}}\delta\br{1-x}
    },
    \\
    P^{(n)}\br{x} &=& \int\limits_x^1 \frac{dy}{y}
    P^{(1)}\br{y} P^{(n-1)}\br{\frac{x}{y}}.\nn
\ea
The structure function $D\br{x,\beta}$ defined in
this way automatically satisfies the Kinoshita-Lee-Nauenberg (KLN)
theorem~\cite{Kinoshita:1962ur,Lee:1964is} on the
cancellation of the mass singularities in the total decay width
\ba
    \int\limits_0^1 dx D\br{x,\beta} = 1.
    \label{KLNdefinition}
\ea
There also exists a smoothed form for the structure function $D\br{x,\beta}$:
\ba
    D\br{z,\beta} =
    2\beta\br{1-z}^{2\beta-1}\br{1+\frac{3}{2}\beta}
    -
    \beta\br{1+z} + O\br{\beta^2},
\ea
which sums radiative corrections in all orders of perturbation theory
which are enhanced by the large logarithmic factor $L$  (in $\beta$) and
is more convenient for numerical evaluation.

The quantity $K\br{x}$ in (\ref{MuonSpecrtumWithRC}) is the so-called $K$-factor
which takes into account the contributions of the radiative corrections
which are not enhanced by the large logarithms and have rather complicated form
(see \cite{Berman:1958ti} or \cite{Berestetsky:1982aq}, \S 147).
We note that, contrary to the singular behaviour ($\sim \ln\br{1-x}$) of $K\br{x}$
in the limit as $x\to 1$, the quantity
\ba
    \int\limits_x^1 \frac{dy}{y}
    D\br{\frac{x}{y},\beta} K\br{y}
\ea
has a finite limit as $x \to 1$ \cite{Bartos:2008kz}.

Thus, applying the general form of the corrected spectrum (\ref{MuonSpecrtumWithRC}),
we obtain the following form of the electron energy spectrum
in the leading logarithmic approximation (LLA):
\ba
    \frac{1}{6\Gamma}
    \frac{d\Gamma}{dx}
    &=&
    2x^2 \brs{1-x - \frac{2}{9} \rho \br{3-4x}}
    +
    \frac{\alpha L}{2\pi} \brs{4F_1\br{x}-\frac{8}{9}\rho F_2\br{x}},
\ea
where the functions $F_{1,2}\br{x}$ are the results of the
application of the structure function to the spectrum in the Born approximation:
\ba
    F_1\br{x} &=& \int\limits_x^1 \frac{dy}{y} y^2\br{1-y} P^{(1)}\br{\frac{x}{y}} =
    \nn\\ &=&
    2x^2\br{1-x}\ln\br{\frac{1-x}{x}} + \frac{1}{6}\br{1-x}\br{1+4x-8x^2}, \\
    F_2\br{x} &=& \int\limits_x^1 \frac{dy}{y} y^2\br{3-4y} P^{(1)}\br{\frac{x}{y}}
    \nn\\ &=&
    2x^2\br{3-4x}\ln\br{\frac{1-x}{x}} + \frac{16}{3}x^3 - 8x^2 + x + \frac{1}{6},
\ea
which satisfy the following property:
\ba
    \int\limits_0^1 dx F_{1,2}\br{x} = 0.
\ea
This is a specific form of the general KLN theorem
\cite{Kinoshita:1962ur,Lee:1964is}.

In concluding this section, we give  the double differential distribution for the case of
the polarised muon decay with the radiative corrections in LLA (here we put $\rho=3/4$):
\ba
\frac{d\Gamma}{\Gamma d x d\cos\theta}\br{\mu\to e\nu_\mu\bar{\nu}_e}&=&
x^2\brs{3-2x-P_\mu (1-2x)\cos\theta}+\nn\\
&+&\frac{\alpha L}{2\pi}\brs{F_3(x)-P_\mu \cos\theta F_4(x)},
\label{BornMuonDecayWithPolarization}
\ea
with $P_\mu$ and  $\theta$ being the degree of muon polarisation and the angle between
the muon polarisation vector and the electron momentum (in the rest frame of the muon).
The functions
\ba
    F_3 = 2\br{6 F_1 - F_2},
    \qquad
    F_4 = 2\br{-2 F_1 + F_2},
\ea
have the explicit expressions:
\ba
F_3(x) &=& 4x^2\br{3-2x}\ln\frac{1-x}{x}+\frac{5}{3}+4x-8x^2+\frac{16}{3}x^3, \nn\\
F_4(x) &=& 4x^2\br{1-2x}\ln\frac{1-x}{x}-\frac{1}{6}-4x^2+8x^3. \nn
\ea

\section{Top quark decays $t\to b (W^+,H^+)$ in the Born approximation}
\label{FirstStepOfDecay}

Top-quark decays within the Standard Model are completely dominated by the
mode
\ba
    t \to b + W^+~, \label{ProcessW}
\ea
due to $V_{tb}=1$ to a very high accuracy.  In beyond-the-SM theories with an extended Higgs
sector, if allowed kinematically, one may also have the decay mode
\ba
    t \to b + H^+ \label{ProcessH}
\ea
where $H^+$ is the charged Higgs boson, which we will consider within the
MSSM. The relevant part of the interaction Lagrangian is \cite{Raychaudhuri:1995kv}:
\ba
    {\cal L}_I &=&
    \frac{g}{2\sqrt{2} M_W} V_{tb}
    H^+ \brs{\bar u_t\br{p_t} \brf{A\br{1+\gamma_5}+B\br{1-\gamma_5}} u_b\br{p_b}} +
    \nn\\&+&
    \frac{g C}{2\sqrt{2} M_W}
    H^+ \brs{\bar u_{\nu_l}\br{p_\nu} \br{1-\gamma_5} u_l\br{p_l}},
    \label{HiggsLagrangian}
\ea
where $A$, $B$ and $C$ are model-dependent parameters which
depend on the fermion masses and $\tan \beta$:
\ba
    A = m_t \cot\beta, \qquad B = m_b \tan \beta, \qquad C = m_\tau \tan \beta.
    \label{LagrangianParametersABC}
\ea
The decay widths of processes (\ref{ProcessW}) and (\ref{ProcessH}) in the Born approximation
are well known \cite{Raychaudhuri:1995kv}:
\ba
    \Gamma_{t \to b W} &=&
    \frac{g^2}{64\pi M_W^2 m_t}
    \lambda^{\frac{1}{2}}\br{1,\frac{m_b^2}{m_t^2},\frac{M_W^2}{m_t^2}}
    \brs{
        M_W^2\br{m_t^2+m_b^2} + \br{m_t^2-m_b^2}^2 - 2M_W^4
    },
    \\
    \Gamma_{t \to b H} &=&
    \frac{g^2}{64\pi M_W^2 m_t}
    \lambda^{\frac{1}{2}}\br{1,\frac{m_b^2}{m_t^2},\frac{M_H^2}{m_t^2}}
    \times\nn\\
    &\times&
    \brs{
        \br{m_t^2\cot^2\beta + m_b^2\tan^2\beta}
        \br{m_t^2 + m_b^2 - M_H^2}
        -
        4 m_t^2 m_b^2
    },
\ea
where $\lambda\br{x,y,z} = x^2+y^2+z^2-2xy-2xz-2yz$ is the
triangle function.
The total top quark decay width then reads as:
\ba
    \Gamma_t^{tot} = \Gamma_{t \to b W} + \Gamma_{t \to b H}.
    \label{TotalWidth}
\ea
We now discuss the total top quark decay width including
the radiative corrections. In the total decay width the
contribution of the QED corrections containing the large logarithms $L$
is cancelled (see (\ref{KLNdefinition})). The non-enhanced
QED corrections are small. The QCD corrections were calculated in
\cite{Czarnecki:1992ig,Czarnecki:1992zm} and have the form:
\ba
    \Gamma_{t,RC}^{tot} &=& \Gamma_{t \to b W}^{Born+QCD} + \Gamma_{t \to b H}^{Born+QCD},
    \label{TotalWidthWithRC}
    \\
    \Gamma_{t \to b (W,H)}^{Born+QCD} &=&
    \Gamma_{t \to b (W,H)}\br{1 + f_{W,H}},
    \qquad
    f_{W,H} =
    \frac{\alpha_s}{3\pi}\br{5-\frac{4\pi^2}{3}}. \nn
\ea

\section{The top quark decay $t\to b W^+ \to b (\ell^+  \nu_\ell)$ in the Born approximation}
\label{TopSection}

The formalism illustrated in Section~\ref{MuonSection} can be used to
discuss the inclusive semileptonic decays of the charm,
beauty and top quarks. However, the decay distributions from the the charm and beauty
hadrons have in addition
important non-perturbative effects, which usually are modelled in terms of the shape functions.
In the case of the top quark decay, since the top quark lifetime is much shorter than the
typical strong interaction time, the decay dynamics is controlled by perturbation theory.
Thus, incorporating the (QED and QCD) perturbative corrections, one has precise theoretical
 predictions
for the energy spectra of the decay products to be confronted with data. We start by
working out the charged lepton energy spectra in the decays
$ t \to b W^+ \to b (\ell^+\nu_\ell)$, where $\ell^+=e^+,\mu^+,\tau^+$. To that end,
let us consider the dominant decay in the SM (see Fig.~\ref{Fig1}, a.)):
\begin{figure}
    \includegraphics[width=0.8\textwidth]{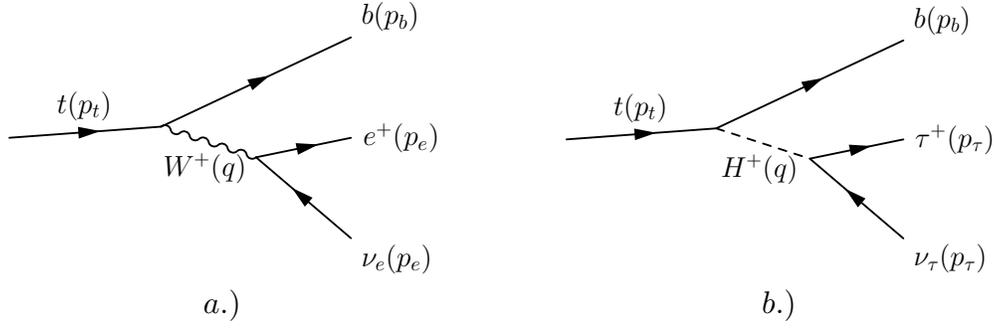}
    \caption{\label{Fig1} Lowest order Feynman diagrams describing the semileptonic
decays of the top quark  a) SM, mediated by $W^+$, b) BSM, mediated by $H^+$.}
\end{figure}
\ba
    t \br{p_t} \to b\br{p_b} + W^+\br{q} \to b\br{p_b} + \br{\ell^+\br{p_\ell} +  \nu_\ell
\br{p_\nu}}~,
\ea
and to be specific, we concentrate on the case with $\ell^+ \nu_\ell =e^+ \nu_e$.
The matrix element of this process in the Born
approximation is given by:
\ba
    M^{t\to b W^+ \to b (e^+ \nu_e)}_{\rm Born} &=&
    i \frac{g^2 V_{tb}}{4\sqrt{2}}\frac{1}{q^2 - M_W^2}
    \br{g_{\mu\nu} - \frac{q_\mu q_\nu}{M_W^2}}\times\nn\\
    &\times&
    \brs{\bar u_b\br{p_b} \gamma^\mu \br{1+\gamma_5} u_t\br{p_t}}
    \brs{\bar u_e\br{p_e} \gamma^\nu \br{1-\gamma_5} u_{\nu_e}\br{p_\nu}},
\ea
where $g^2 = \frac{4\pi \alpha}{\sin^2\theta_W} = 8\pi M_W^2 G_F/\sqrt{2}$
is the electroweak coupling constant,
$\theta_W$ is the weak mixing angle, $M_W$ is the $W^\pm$-boson mass,
 and $V_{tb}$ is an
element of the Cabibbo-Kobayashi-Maskawa (CKM) quark mixing
 matrix~\cite{Cabibbo:1963yz,Kobayashi:1973fv}.
We note that the contribution of the second term in the parenthesis is proportional to the
electron mass due to the conservation of the lepton current and can be omitted.
The matrix element squared then reads as:
\ba
    \brm{M^{t\to b W^+ \to b (e^+  \nu_e)}_{\rm Born}}^2 =
    \br{\frac{g^2 V_{tb}}{4\sqrt{2}\br{q^2 - M_W^2}}}^2
    2^8 \br{p_b p_\nu}\br{p_t p_e}.
    \label{WDecayMatrixSquare}
\ea

Let us introduce the following notation for the kinematic variables:
\ba
    \begin{array}{lclcl}
        x_b=\frac{2 E_b}{m_t}, &\qquad& x_e=\frac{2E_e}{m_t}, &\qquad& x_\nu=\frac{2E_\nu}{m_t}, \\
        \gamma = \frac{\Gamma_W}{M_W}, &\qquad& \xi = \frac{m_t^2}{M_W^2}, &\qquad& \eta = \frac{m_b^2}{m_t^2}, \\
    \end{array}
    \nn
\ea
where $E_b$, $E_e$ and $E_\nu$ are the energies of the $b$-quark, positron and neutrino in
the $t$-quark rest frame, respectively. $\Gamma_W$ is the total decay width of the $W$-boson.
In terms of these variables, the various scalar products can be expressed as:
\ba
    \begin{array}{lclcl}
        2\br{p_b p_e} = m_t^2\br{1-x_\nu-\eta}, &\quad&   2\br{p_b p_\nu} = m_t^2\br{1-x_e-\eta},
 &\quad& 2\br{p_e p_\nu} = m_t^2\br{1+\eta-x_b},\\
        2\br{p_t p_e} = m_t^2 x_e, &\quad& 2\br{p_t p_\nu} = m_t^2 x_\nu, &\quad& 2\br{p_t p_b} =
 m_t^2 x_b.\\
    \end{array}
    \nn
\ea
Since the main contribution to this decay comes from the kinematic region
where $W$-boson is near its mass-shell we have to take into account its decay width.
We use the Breit-Wigner form of the propagator:
\ba
    \frac{1}{\brm{q^2 - M_W^2}^2} \to
    \frac{1}{\brm{q^2 - M_W^2 + i M_W \Gamma_W}^2} =
    \frac{1}{M_W^4} \frac{1}{\br{1-\xi\br{1+\eta-x_b}}^2 + \gamma^2}.
\ea

Thus, the matrix element squared (\ref{WDecayMatrixSquare}) then reads
\ba
    \brm{M^{t\to b W^+ \to b (e^+  \nu_e)}_{\rm Born}}^2 =
    \frac{2 \br{g^2 V_{tb}}^2 x_e \br{1-x_e-\eta}}{\br{1-\xi\br{1+\eta-x_b}}^2 + \gamma^2}
    \xi^2.
\ea

The phase space  volume element with three-particle final state has the standard form:
\ba
    d\Phi_3 =
    \br{2\pi}^{-5} \delta\br{p_t-p_b-p_e-p_\nu}
    \frac{d \vec p_b}{2E_b} \frac{d \vec p_e}{2E_e} \frac{d \vec p_\nu}{2E_\nu}
    =
    \frac{m_t^2}{2^7 \pi^3} dx_e dx_\nu.
\ea
The kinematic restrictions are:
\ba
    0 \le &x_e& \le 1-\eta, \nn\\
    1-x_\tau \le &x_b& \le 1, \nn\\
    1-x_e-\eta \le &x_\nu& \le 1-\frac{\eta}{1-x_e}, \nn
\ea
and the $b$-quark mass-shell condition fixes the cosine of the angle between the positron and
the neutrino momenta directions $C_{e\nu}=\cos\br{\theta_{e\nu}} =\frac{ \vec{p}_e.
\vec{p}_\nu}{\brm{\vec{p}_e}\brm{\vec{p}_\nu}}$,
%
\ba
    C_{e\nu} = 1 + \frac{2}{x_e x_\nu}\br{1-x_e-x_\nu-\eta}.
\ea

On using the standard formulae for the decay width
\ba
    d\Gamma = \frac{1}{2 \cdot 2m_t} \brm{M}^2 d\Phi_3,
\ea
we obtain for the case of the unpolarised top quark decay $t\to b W^+ \to b (\ell^+  \nu_\ell)$
the decay width:
\ba
    \frac{d\Gamma^{t\to b W^+ \to b (l^+  \nu_l)}_{\rm Born}}{dx_b dx_l} &=&
    \Gamma_t
    \frac{x_l \br{1-x_l-\eta}}{\br{1-\xi\br{1+\eta-x_b}}^2 + \gamma^2}
    =
    \Gamma_t
    \frac{x_l\br{x_l^{max}-x_l}}{\br{1-\frac{y}{y_0}}^2+\gamma^2},
    \label{BornDecayWithW}
\ea
where $y=1+\eta-x_b$, $x_e^{max}=1-\eta$ and $y_0 = 1/\xi$. $\Gamma_t$ is the
dimensional factor:
\ba
    \Gamma_t&=&\frac{G_F^2 m_t^5 V_{tb}^2}{16\pi^3}.
    \label{GammaTW}
\ea

Now, we calculate the branching ratios of the decays considered above. The
branching ratio of the decay $t\to b W^+ \to b (\ell^+  \nu_\ell)$ is obtained from
(\ref{BornDecayWithW}) by dividing it by the total width of top quark
$\Gamma_t^{tot}$ (see (\ref{TotalWidth})):
\ba
    \frac{dBr^{t\to b W^+ \to b (l^+  \nu_l)}_{\rm Born}}{dx_b dx_l} &=&
    B_t
    \frac{x_l\br{x_l^{max}-x_l}}{\br{1-\frac{y}{y_0}}^2+\gamma^2},
    \qquad
    B_t = \frac{\Gamma_t}{\Gamma_t^{tot}},
    \label{BrBornDecayWithW}
\ea
Let us consider the electron energy spectrum. In the Born approximations it has the
following expression:
\ba
    \frac{dBr^{(0)}}{dx_e}
    &\equiv&
    \frac{dBr^{t\to b W^+ \to b (e^+ \nu_e)}_{\rm Born}}{dx_e}
    =
    \int\limits_{1-x_e}^{1}
    dx_b
    \frac{dBr^{t\to b W^+ \to b (e^+ \bar \nu_e)}_B}{dx_b dx_e}
    =\nn\\
    &=&
    B_t \cdot x_e\br{x_e^{max}-x_e} \Phi_W\br{x_e},
    \label{ElectronSpectrumBorn}
\ea
where
\ba
    \Phi_W\br{x}
    &=&
    \int\limits_0^x \frac{dy}{\br{1-\frac{y}{y_0}}^2+\gamma^2} =
    \nn\\
    &=&
    \frac{1}{\gamma\xi}
    \brs{
        \arctan\br{\frac{\xi\br{1-\sqrt{\eta}}^2-1}{\gamma}} +
        \arctan\br{\frac{\xi\br{\eta+x}-1}{\gamma}}
    }.
    \label{PhiPrecise}
\ea

\subsection{QCD radiative corrections}

The inclusive electron energy spectrum including  the lowest order
 QCD corrections is
\ba
    \frac{dBr^{t\to b W^+ \to b (e^+ \nu_e)}_{Born+QCD}}{dx_e}
    =
    \frac{1}{\Gamma_{t,RC}^{tot}}
    \br{
        \frac{d\Gamma^{(0)}}{dx_e}
        +
        \frac{d\Gamma^{(1)}_{QCD}}{dx_e}
    },
    \label{WQCDCorr}
\ea
where $\Gamma_{t,RC}^{tot}$ is the radiatively corrected
total decay width of top quark from (\ref{TotalWidthWithRC}).
This expression is free from the $b$-quark mass singularities, hence
 we can put $\eta = 0$, which yields:
\ba
    \frac{d\Gamma^{(1)}_{QCD}}{dx_e}
    =
    -\Gamma_t \frac{2\alpha_s}{3\pi}
    \int\limits_0^{x_e}
    \frac{dy}{\br{1-\xi y}^2 + \gamma^2}
    F_W\br{x_e,y},
    \label{QCDCorrections}
\ea
where the function $F_W\br{x,y}$ is finite in the limit $m_b\to 0$ and has the form
\cite{Jezabek:1988ja}:
\ba
    F_W\br{x,y} &=&
    2x\br{1-x}\brs{ \zeta_2 + \Li{2}\br{x} + \Li{2}\br{\frac{y}{x}} +
    \frac{1}{2}\ln^2\br{\frac{1-y/x}{1-x}}}
    +\nn\\&+&
    x \brs{ \zeta_2 + \Li{2}\br{y} - \Li{2}\br{x} - \Li{2}\br{\frac{y}{x}} }
    +\nn\\&+&
    \frac{1}{2} \ln\br{1-y} \brs{ -\br{3+2x} + 2y\br{1+x} + y^2 }
    +\nn\\&+&
    \frac{1}{2} \ln\br{1-\frac{y}{x}} \brs{ x\br{9-4x} - 2y\br{1+x} - y^2 }
    +\nn\\&+&
    \frac{5\br{1-x}}{2} \ln\br{1-x} + \frac{1}{2} y \br{1-x} \br{\frac{y}{x}+4}.
    \label{Fdef}
\ea
This formula is valid for $x_e < 1$. For  $x_e \approx 1$, close to the boundary
of the phase space, there are Sudakov Logarithms due to the limited phase space,
and this result becomes unstable but remains integrable.
The electron energy spectrum with the QCD corrections is given by:
\ba
    \frac{dBr^{t\to b W^+ \to b (e^+ \nu_e)}_{Born+QCD}}{dx_e}
    =
    \frac{\Gamma_t}{\Gamma_{t,RC}^{tot}}
    \int\limits_0^{x_e}
    \frac{dy}{\br{1-\xi y}^2 + \gamma^2}
    \brs{ x_e\br{x_e^{max}-x_e} - \frac{2\alpha_s}{3\pi} F_W\br{x_e,y} }.
    \label{ElectronSpectrumBornQCD}
\ea
%
\subsection{QED radiative corrections in the leading logarithmic approximation}

\begin{figure}
    \includegraphics[width=0.8\textwidth]{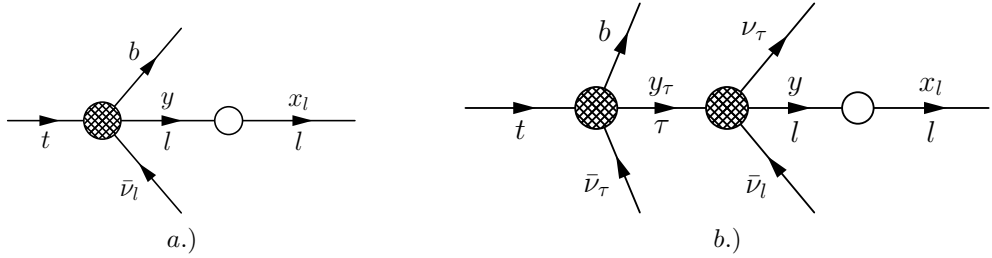}
    \caption{\label{FigSFKinematicsLeptons}
    Kinematics depicting the application of the structure function
    method, which involves factorisation of the amplitude in the
     "hard sub-process" (filled circle) and the
    "long-distance" contributions (empty circle) taken into
    account by the convolution with the structure function $D_l(x,\beta)$
    (see (\ref{ElectronSpectrumRC}).}
\end{figure}

To calculate the QED radiative corrections in the leading logarithmic approximation
we will use the SF method which was illustrated in Section~\ref{MuonSection}
(see kinematic scheme in Fig.~\ref{FigSFKinematicsLeptons}, a).

The QED radiative corrected spectrum is:
\ba
    \frac{dBr^{t\to b W^+ \to b (e^+ \bar \nu_e)}_{Born+QED}}{dx_e}
    =
    \frac{1}{\Gamma_{t,RC}^{tot}}
    \int\limits_{x_e}^1 \frac{dy_e}{y_e} D\br{\frac{x}{y_e},\beta_e}
    \frac{d\Gamma^{t\to b W^+ \to b (e^+ \bar \nu_e)}_{B}}{dy_e},
    \label{ElectronSpectrumRC}
\ea
where the structure function $D\br{x,\beta_e}$ was defined in (\ref{SF}).

\begin{figure}
    \includegraphics[width=0.8\textwidth]{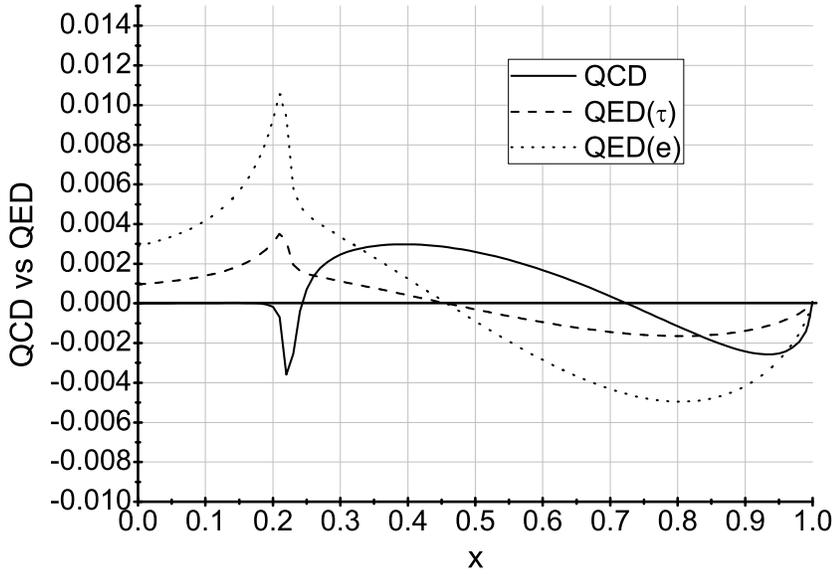}
    \caption{\label{FigQCDvsQED}
    QCD and QED corrections to the lepton energy spectrum in the decays
    $t \to b W^+ \to b (e^+\nu_e, \tau^+\nu_\tau)$.
    The solid curve is the QCD-correction term $\frac{dBr^{(1)}_{QCD}}{dx_e}$
    (i.e. second term from from (\ref{ElectronSpectrumBornQCD})),
    the dashed and dotted curves are the QED-corrections
    $\frac{dBr^{t\to b W^+ \to b (\tau^+ \nu_\tau, e^+ \nu_e)}_{QED_{LLA}}}{dx_{\tau,e}}$
    from (\ref{QEDCorrection}) for the $\tau^+$ and $e^+$ in the final state,
respectively
    }
\end{figure}

The first order QED radiative correction reads as (using the
electron energy spectrum in the Born approximation (\ref{ElectronSpectrumBorn})):
\ba
    \frac{dBr^{t\to b W^+ \to b (e^+ \bar \nu_e)}_{QED_{LLA}}}{dx_e}
    &=&
    \frac{\alpha}{2\pi}\br{L_e-1}
    \frac{1}{\Gamma_{t,RC}^{tot}}
    \int\limits_{x_e}^1 \frac{dy_e}{y_e} P^{(1)}\br{\frac{x_e}{y_e}}
    \frac{d\Gamma^{t\to b W^+ \to b (e^+ \bar \nu_e)}_{B}}{dy_e} \nn\\
    &=&
    \frac{\alpha}{2\pi}\br{L_e-1}\frac{\Gamma_t}{\Gamma_{t,RC}^{tot}}\int\limits_{x_e}^1
    \frac{dy_e}{y_e} P^{(1)}\br{\frac{x_e}{y_e}}
    y_e\br{y_e^{max}-y_e} \Phi_W\br{y_e} \nn\\
    &=&
    \frac{\alpha}{2\pi}\br{L_e-1}\frac{\Gamma_t}{\Gamma_{t,RC}^{tot}}~I\br{x_e},
    \label{QEDCorrection}
\ea
where $y_e^{max}=1-\eta$, and
\ba
    I\br{x} &=&
    \int\limits_{x}^1 \frac{dy}{y} P^{(1)}\br{\frac{x}{y}}
    y\br{y^{max}-y} \Phi_W\br{y} = \nn\\
    &=&
    \Phi_W\br{x}
    \brf{
        x\br{1-x} \brs{2\ln\br{\frac{1-x}{x}}+\frac{3}{2}}
        + x\ln\br{x} + \br{1-x}^2 - \frac{1}{2}\br{1-x^2}
    }
    +\nn\\
    &+&
    \int\limits_x^1
    dy \frac{\br{1-y}\br{y^2+x^2}}{y\br{y-x}}
    \brs{\Phi_W\br{y}-\Phi_W\br{x}},
    \label{IPrecise}
\ea
where $\Phi_W\br{x}$ is given in (\ref{PhiPrecise}).
The contribution of the QCD correction $ \frac{dBr^{(1)}_{QCD}}{dx_e}$
from (\ref{WQCDCorr}) is shown in Fig.~\ref{FigQCDvsQED}, and is the same
for $\ell =e, \mu, \tau$. The contributions of the QED corrections is specific
to the charged lepton $e, \mu, \tau$ and shown in Fig.~\ref{FigQCDvsQED}.
The input parameters used in this figure and subsequently are given in a
 table in
 the Appendix. In Fig.~\ref{FigElectronSpectrum}, we show the electron energy
 spectrum in the Born approximation and compare it with the (QED + QCD) radiatively
 corrected ones.

It is obvious from the foregoing that the QED radiative corrections break the
lepton universality, encoded at the Lagrangian level for the decays
$t \to b W^+ \to b \ell^+ \nu_\ell$. It is also clear that the radiative
corrections are not overall multiplicative renormalizations and they distort
the Born level distributions in a non-trivial way.
 To quantify this, we plot the ratios
$R_{e \tau} (x)$ and $R_{\mu \tau}(x)$, defined below, in Fig.~\ref{FigRatio}.

\ba
R_{e\tau(x)} &=& \frac{\left(\Gamma^{t \to bW^+ \to b (\tau^+ \nu_\tau)}
 (x=x_\tau)\right)_{\rm Born + QCD + QED_{LLA}}}
{\left(\Gamma^{t \to bW^+ \to b (e^+ \nu_e)}
 (x=x_e)\right)_{\rm Born + QCD + QED_{LLA}} } \nn \\
& R_{\mu\tau(x)}=& \frac{\left(\Gamma^{t \to bW^+ \to b (\tau^+ \nu_\tau)}
 (x=x_\tau)\right)_{\rm Born + QCD + QED_{LLA}}}
{\left(\Gamma^{t \to bW^+ \to b (\mu^+ \nu_e)}
 (x=x_\mu)\right)_{\rm Born + QCD + QED_{LLA}} }
\label{Defratio}
\ea

As can be seen, the effect of the radiative corrections is very marked for the
low-$x$ values of the lepton-energy spectra ($x \leq 0.3)$ and it is non-neglible
also  near the end-point of the spectra ($  x \geq 0.7$).
This is numerically an important effect and in the precision tests of the SM in
the top-quark sector, which we anticipate will be carried out at the LHC,
it is mandatory to take the radiative distortions of the spectra into account.

\begin{figure}
    \includegraphics[width=0.8\textwidth]{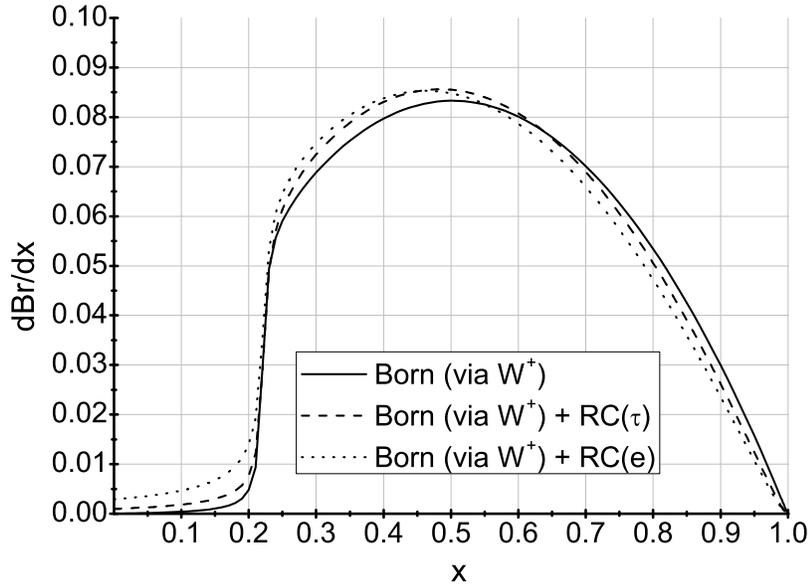}
    \caption{\label{FigElectronSpectrum}
    Lepton energy spectra from the decays $t \to b W^+ \to b (\ell^+ \nu_\ell)$
    versus the lepton energy fraction $x=x_{e,\mu,\tau}$.
    The spectrum in the Born approximation (solid curve) is the same for
    $\ell^+=e^+,\mu^+,\tau^+$ (see (\ref{BornDecayWithW})).
    The dotted curve is the $e^+$-energy spectrum ($dBr/dx_e$)
    including the (QCD+QED) radiative corrections for the decay
    $t \to b W^+ \to b\br{e\nu_e}$
    (i.e. the contributions from (\ref{ElectronSpectrumBornQCD})
    plus the QED correction term from (\ref{QEDCorrection})).
    The dashed curve is the $\tau^+$-energy spectrum ($dBr/dx_\tau$)
    including the (QCD+QED) radiative corrections for the decay
    $t \to b W^+ \to b\br{\tau\nu_\tau}$.
}
\end{figure}

\begin{figure}
    \includegraphics[width=0.8\textwidth]{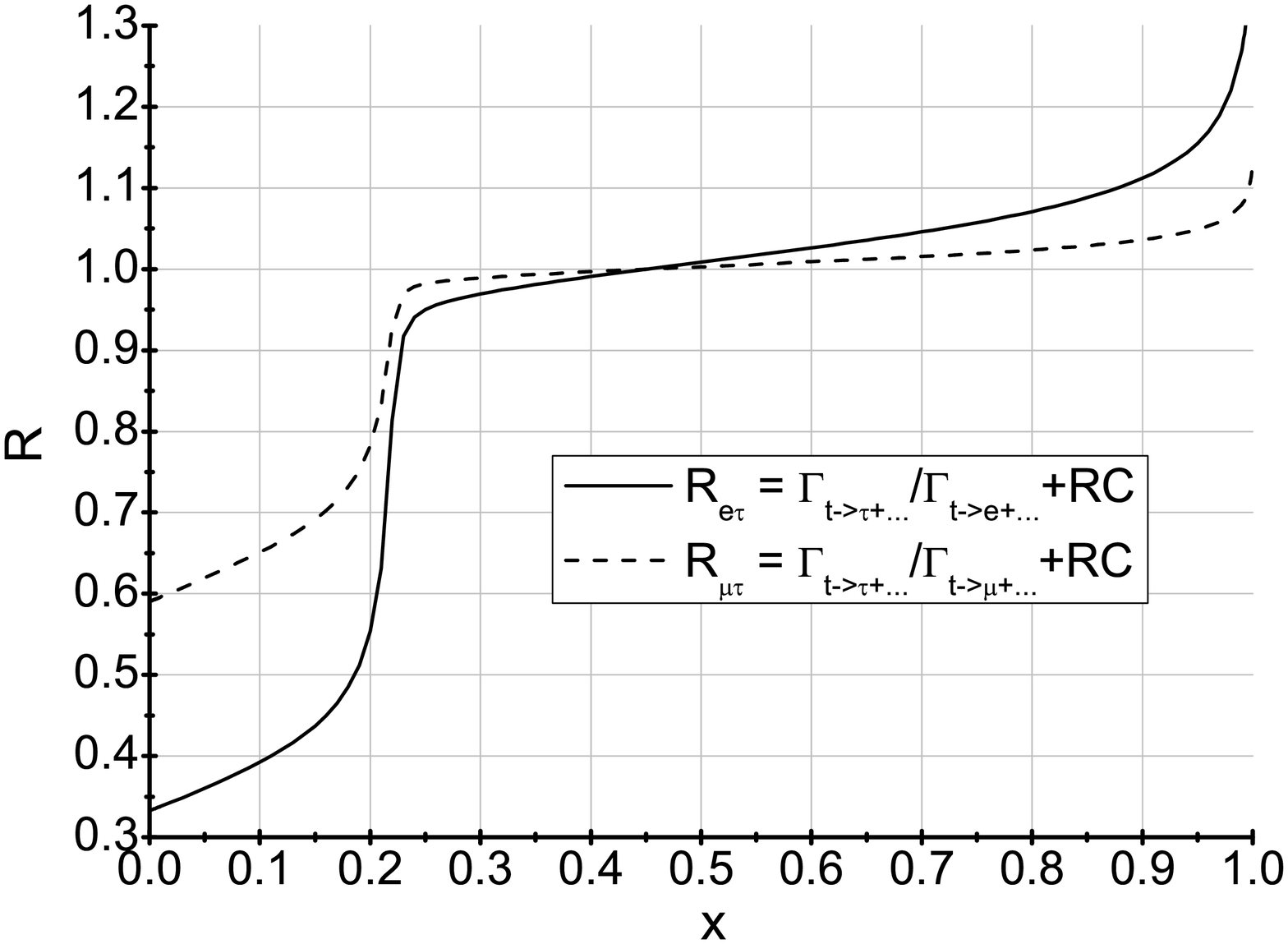}
    \caption{\label{FigRatio}
    Ratios of the lepton-energy spectra  in the decays
 $t \to b W^+ \to b (\ell^+\nu_\ell)$:
        $R_{e\tau}(x) = \frac{d\Gamma^{t \to b W^+ \to b\br{\tau\nu_\tau}}}{dx_\tau}/
                    \frac{d\Gamma^{t \to b W^+ \to b\br{e\nu_e}}}{dx_e}$
  (solid curve), and
       $R_{\mu\tau}(x) = \frac{d\Gamma^{t \to b W^+ \to b\br{\tau\nu_\tau}}}
 {dx_\tau}/ \frac{d\Gamma^{t \to b W^+ \to b\br{\mu\nu_\mu}}}{dx_\mu}$
(dashed curve),  quantifying the leading order (QCD and QED) corrections to
 the lepton universality in semileptonic top quark decays.
}
\end{figure}

\section{The top quark decay $t\to b H^+ \to b (l^+ \bar \nu_l)$ in the Born approximation}
\label{TopHiggsDecays}

Let us consider now the top quark decay induced by a charged Higgs boson:
\ba
    t \br{p_t} \to b\br{p_b} + H^+\br{q} \to b\br{p_b} + \br{\ell^+\br{p_\ell} + \bar \nu_\ell
\br{p_\nu}}
    \label{ProcessWithHiggs}
\ea
where we will concentrate on $\ell^+=\tau^+$ (see Fig.~\ref{Fig1}, b).
Using the couplings from the Lagrangian (\ref{HiggsLagrangian}) we can write the
matrix element of the process (\ref{ProcessWithHiggs}) in the following form
\ba
    M^{t\to b H^+ \to b (\tau^+  \nu_l)}_B &=&
    i \frac{g^2 V_{tb}}{8 M_W^2}\frac{C}{q^2 - M_H^2 + i M_H \Gamma_H}
    \brs{\bar u_{\nu_\tau}\br{p_\nu} \br{1+\gamma_5} u_\tau\br{p_\tau}}
    \times\nn\\
    &\times&
    \brs{\bar u_t\br{p_t} \brf{A\br{1-\gamma_5}+B\br{1+\gamma_5}} u_b\br{p_b}}.
\ea
The model parameters $A$, $B$, $C$ are given in (\ref{LagrangianParametersABC}).
Squaring this matrix element yields

\ba
    \brm{M^{t\to b H^+ \to b (\tau^+  \nu_\tau)}_B}^2 =
    \frac{\br{g^2 V_{tb}}^2}{\br{q^2-M_H^2}^2+M_H^2 \Gamma_H^2}
    \br{p_t p_b}\br{p_\tau p_\nu}
    \frac{C^2 \br{A^2+B^2}}{M_W^4}.
    \label{HDecayMatrixSquare}
\ea

\begin{figure}
    \includegraphics[width=0.8\textwidth]{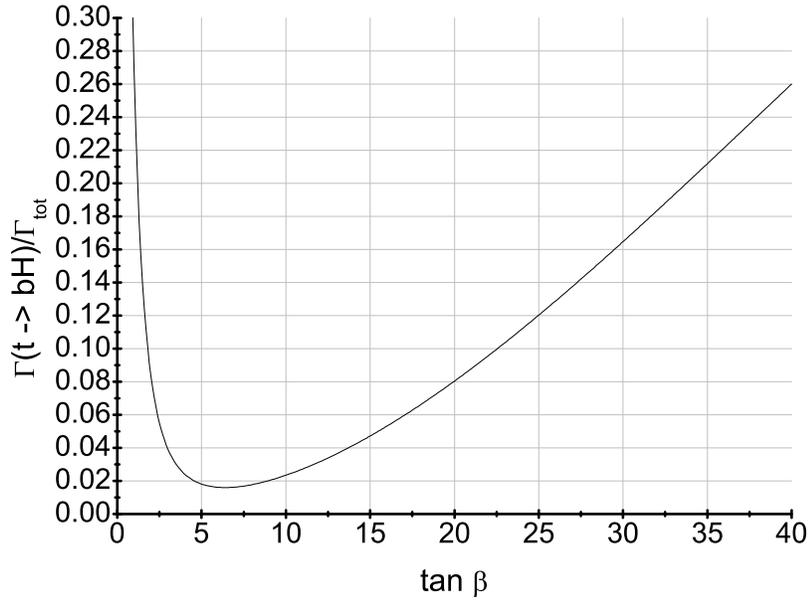}
    \caption{\label{FigTanBetaDependence}
    Lowest order branching ratio for the decay $t\to b H^+$ as a function of $\tan\beta$
for $M_{H^+}=120$ GeV.
    }
\end{figure}

Introducing the kinematic variables:
\ba
    \begin{array}{lclcl}
        x_\tau=\frac{2E_\tau}{m_t}, &\qquad& x_\nu=\frac{2E_{\nu_\tau}}{m_t}, &\qquad& y=\frac{q^2}{m_t^2}=1+\eta-x_b, \\
        \gamma_H = \frac{\Gamma_H}{M_H}, &\qquad& y_0 = \frac{M_H^2}{m_t^2}, &\qquad& x_b^{max}=1+\eta,  \\
    \end{array}
    \nn
\ea
where $E_\tau$ and $E_{\nu_\tau}$ are the energies of the final $\tau$ lepton and the neutrino in
the $t$-quark rest frame, respectively. $\Gamma_H$ is the total decay width of the charged
 Higgs boson, and the Breit-Wigner form of the propagator reads as
\ba
    \frac{1}{\brm{q^2 - M_H^2 + i M_H \Gamma_H}^2} =
    \frac{1}{\br{q^2-M_H^2}^2+M_H^2 \Gamma_H^2} =
    \frac{1}{M_H^4} \frac{1}{\br{1-\frac{y}{y_0}}^2 + \gamma_H^2}.
\ea
Thus the matrix element squared (\ref{HDecayMatrixSquare}) takes the form:
\ba
    \brm{M^{t\to b H^+ \to b (\tau^+  \nu_\tau)}_B}^2 =
    \br{g^2 V_{tb}}^2\frac{C^2 \br{A^2+B^2}}{M_W^4}
    \frac{x_b\br{x^{max}-x_b}}{\br{1-\frac{y}{y_0}}^2 + \gamma_H^2}
    \frac{1}{4 y_0^2}.
    \label{HDecayMatrixSquareSimplified}
\ea
The decay width of the unpolarised top quark decay $t\to b H^+ \to b (\tau^+ \bar \nu_\tau)$
then takes the form:
\ba
    \frac{d\Gamma^{t\to b H^+ \to b (\tau^+  \nu_\tau)}_B}{dx_b dx_\tau} &=&
    \Gamma_t^H
    \frac{x_b\br{x^{max}-x_b}}{\br{1-\frac{y}{y_0}}^2+\gamma_H^2},
    \label{BornDecayWithH}\\
    \Gamma_t^H &=& \frac{1}{2^{11}\pi^3} \br{\frac{C^2\br{A^2+B^2}}{M_W^4}}\br{g V_{tb}}^2
    \frac{m_t^5}{M_H^4}
    .
    \label{GammaTH}
\ea
The branching ratio of this decay is:
\ba
    \frac{dBr^{t\to b H^+ \to b (\tau^+  \nu_\tau)}_B}{dx_b dx_\tau} &=&
    B_t^H B_\tau^H
    \frac{x_b\br{x^{max}-x_b}}{\br{1-\frac{y}{y_0}}^2+\gamma_H^2},
    \qquad
    B_t^H = \frac{\Gamma_t^H}{\Gamma_t^{tot}},
    \label{BrTH}
\ea
where $B_\tau^H$ is the branching of the decay
 $H^+ \to \tau^+\nu_\tau$ \cite{Raychaudhuri:1995kv}:
\ba
    B_\tau^H &=& \frac{\Gamma_{H \to \tau \nu_\tau}}{\Gamma_{H \to \tau \nu_\tau}+\Gamma_{H \to c\bar s}}~, \\
    \Gamma_{H \to \tau \nu_\tau} &=&
    \frac{g^2 M_H}{32 \pi M_W^2} m_\tau^2 \tan^2 \beta, \nn\\
    \Gamma_{H \to c\bar s} &=&
    \frac{3 g^2 M_H}{32 \pi M_W^2} \br{M_C^2 \cot^2 \beta + M_S^2 \tan^2\beta}. \nn
\ea
For the numerical values of $\tan \beta$ that we entertain in this paper,
the branching ratio  $B_\tau^H =1$, to a very high accuracy.
The dependence of the branching ratio of the decay $t\to b H^+$ on
 $\tan\beta$ is plotted in Fig.~\ref{FigTanBetaDependence}. We emphasize that
 in plotting this figure, radiative corrections coming from the supersymmetric
 sector are not included. They have been calculated in great detail in the
literature, in particular
 for the MSSM scenario in \cite{Carena:1999py}, and can be effectively incorporated
by replacing the $b$-quark mass $m_b$ in the Lagrangian for the decay
 $t \to  b H^+$ by the SUSY-corrected mass
 $m_b^{\rm corrected}=m_b/[1 + \Delta_b]$. The correction $\Delta_b$ is a
 function of the supersymmetric parameters and, for given MSSM scenarios,
this can be calculated using the FeynHiggs programme~\cite{Hahn:2008zzd},
which makes use of the results in \cite{Carena:1999py}.
In particular, for large values of $\tan \beta$ (say, $\tan \beta > 20)$),
the MSSM corrections increase the branching ratio for $t \to bH^+$ significantly
though this is numerically not important for $\tan \beta=22$, which we use to
numerically calculate the branching ratio for $t \to b H^+$. We emphasize
that in the analysis of data in the MSSM context, the branching ratio
shown here in Fig.~\ref{FigTanBetaDependence} has to be corrected to include
the SUSY corrections. This, for example, can be seen in a
particular MSSM scenario in a recent update~\cite{Sopczak:2009sm}, based
on the version FeynHiggs v2.6.2.

The lepton energy spectrum in the Born approximations has the
following expression
\ba
    \frac{dBr^{t\to b H^+ \to b (\tau^+ \bar \nu_\tau)}_B}{dx_\tau}
    =
    \int\limits_{1-x_\tau}^{1}
    dx_b
    \frac{dBr^{t\to b H^+ \to b (\tau^+  \nu_\tau)}_B}{dx_b dx_\tau}
    =
    B_t^H B_\tau^H \cdot \Phi_H\br{x_\tau},
    \label{TauSpectrumBorn}
\ea
where
\ba
    \Phi_H\br{x} &=&
    \int\limits_0^x dy
    \frac{y\br{1-y}}{\br{1-\frac{y}{y_0}}^2+\gamma^2}
    \nn\\
    &=& y_0^2
    \left\{
        \frac{1-y_0}{\gamma_H}
        \brs{\arctan\br{\frac{1}{\gamma_H}} + \arctan\br{\frac{x-y_0}{y_0 \gamma}}}
    \right.+\nn\\
    &+&\left.
        \br{2y_0-1}\ln\br{\frac{y_0}{y_0-x}} - x
    \right\}.
    \label{PhiHPrecise}
\ea

\subsection{QCD radiative corrections}

The leading order QCD corrections to the decay $t\to b H^+ \to b (\tau^+  \nu_\tau)$
is calculated in a similar way as for the case of $t\to b W^+ \to b (\tau^+  \nu_\tau)$.
The derivations for the Dalitz distribution $dBr/dx_b dx_\tau$ and the $\tau$-energy
spectrum $dBr/dx_\tau$ are given in Appendix~\ref{RCToHiggs}.

\subsection{QED radiative corrections in the leading logarithmic approximation}

\begin{figure}
    \includegraphics[width=0.8\textwidth]{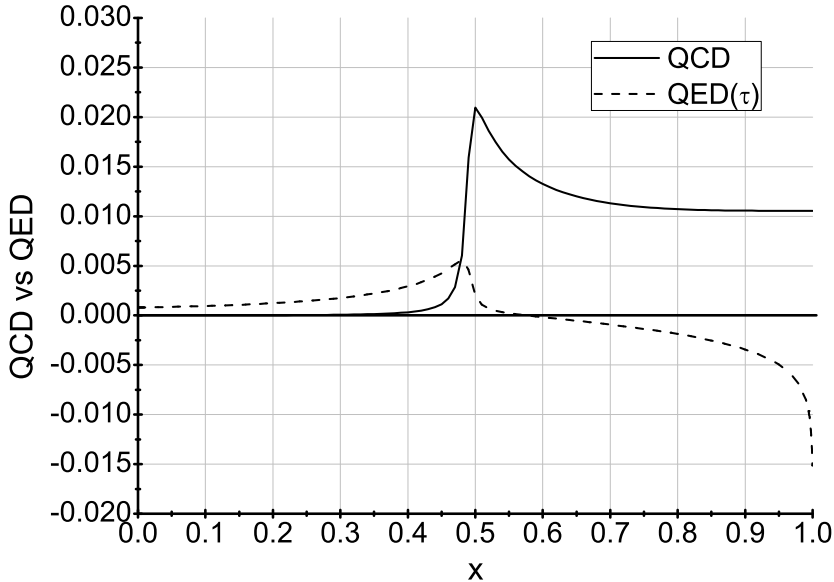}
    \caption{\label{FigHQCDvsQED}
    QCD and QED corrections to the lepton energy spectrum in the decays
    $t \to b H^+ \to b \tau^+\nu_\tau$.
    The solid curve is the QCD corrections, i.e. the second term on the
    r.h.s. of the first line
    of Eq.~(\ref{DefFH}) divided by the total decay width $\Gamma_{t,RC}^{tot}$,
    and the dashed curve is the QED corrections
    $\frac{dBr^{t\to b H^+ \to b \tau^+ \nu_\tau}_{QED_{LLA}}}{dx_{\tau}}$
    from Eq.~(\ref{HQEDCorrection}) for the $\tau^+$ in the final state.
    }
\end{figure}

\begin{figure}
    \includegraphics[width=0.8\textwidth]{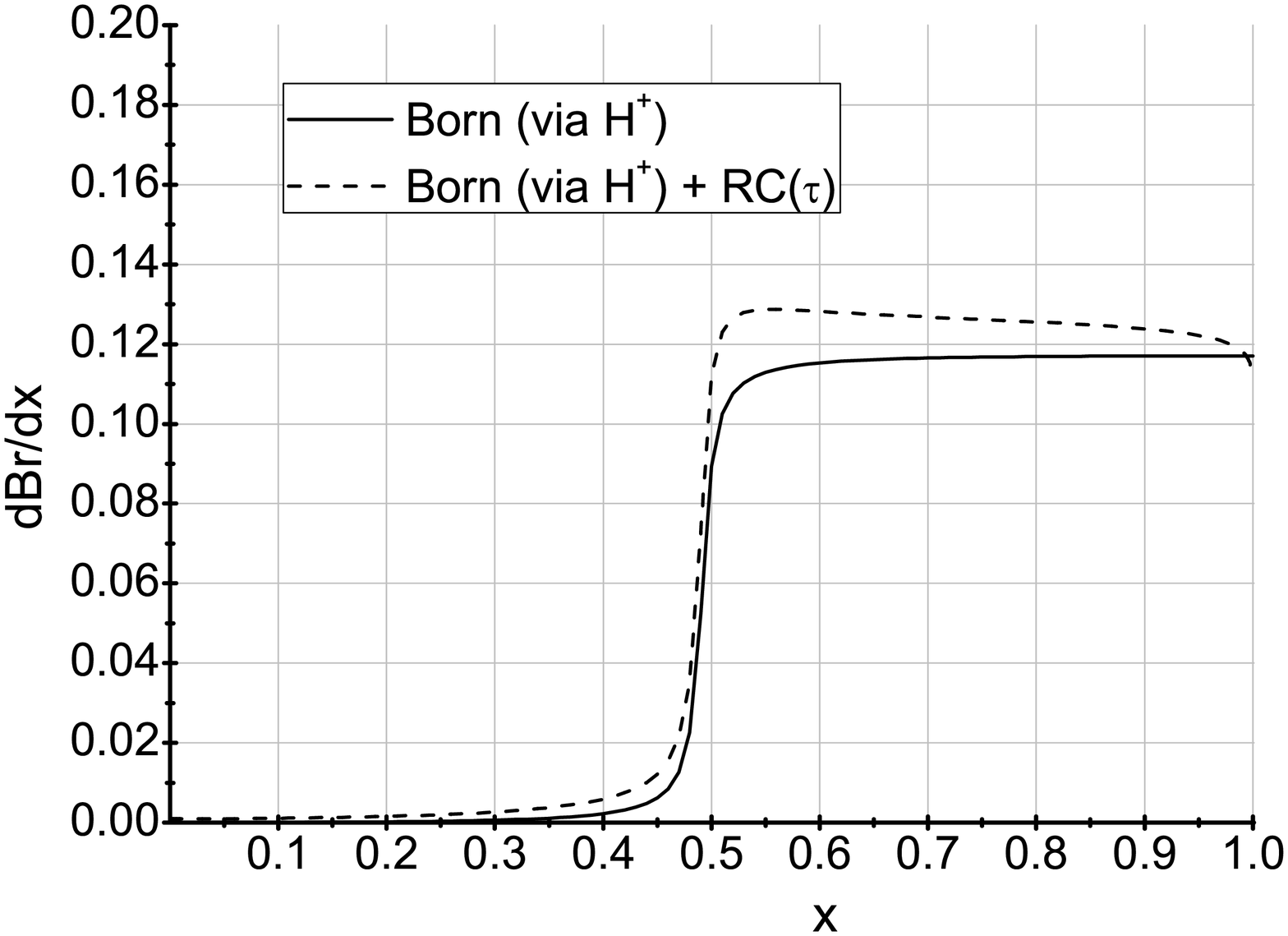}
    \caption{\label{FigTauSpectrum}
    Lepton energy spectra from the decays $t \to b H^+ \to b (\tau^+ \nu_\tau)$
    versus the $\tau$ energy fraction $x=x_\tau$ for $M_{H^+}=120$ GeV and
$\tan \beta=22$. The solid curve shows the Born spectrum
    (see (\ref{TauSpectrumBorn})) and the dashed curve is the spectrum
    including the (QED + QCD) radiative corrections from Fig.~\ref{FigHQCDvsQED}
    (i.e. QCD corrections are taken from Eq.~(\ref{DefFH}) and
    the QED corrections are taken from (\ref{HQEDCorrection})).
    }
\end{figure}

To calculate the QED radiative corrections in the leading logarithmic approximation
we will again use the structure function method, which gives:
\ba
    \frac{dBr^{t\to b H^+ \to b (\tau^+  \nu_\tau)}_{Born+QED}}{dx_\tau}
    =
    \frac{B_\tau^H}{\Gamma_{t,RC}^{tot}}
    \int\limits_{x_\tau}^1 \frac{dy}{y} D\br{\frac{x_\tau}{y},\beta_\tau}
    \frac{d\Gamma^{t\to b H^+ \to b (\tau^+  \nu_\tau)}_{B}}{dy}.
    \label{TauSpectrumRC}
\ea
where the large QED logarithm now is $\beta_\tau$ (see (\ref{FigQCDvsQED})).
The first order QED radiative correction reads as
\ba
    \frac{dBr^{t\to b H^+ \to b (l^+  \nu_l)}_{QED_{LL}}}{dx_l}
    &=&
    \frac{\alpha}{2\pi}\br{L_l-1}
    \frac{B_\tau^H}{\Gamma_{t,RC}^{tot}}
    \int\limits_{x_l}^1 \frac{dy}{y} P^{(1)}\br{\frac{x_l}{y}}
    \frac{d\Gamma^{t\to b H^+ \to b (l^+  \nu_l)}_{B}}{dy} \nn\\
    &=&
    \frac{\alpha}{2\pi}\br{L_l-1}
    \frac{\Gamma_t^H}{\Gamma_{t,RC}^{tot}} B_\tau^H
    \int\limits_{x_l}^1 \frac{dy}{y} P^{(1)}\br{\frac{x_l}{y}}
    \Phi_H\br{y_l} \nn\\
    &=&
    \frac{\alpha}{2\pi}\br{L_l-1}
    \frac{\Gamma_t^H}{\Gamma_{t,RC}^{tot}} B_\tau^H
    I_H\br{x_l},
    \label{HQEDCorrection}
\ea
where $\Phi_H\br{x}$ is the Born spectrum defined in (\ref{PhiHPrecise}) and
\ba
    I_H\br{x} &=&
    \int\limits_{x}^1 \frac{dy}{y} P^{(1)}\br{\frac{x}{y}} \Phi_H\br{y} \nn\\
    &=&
    \Phi_H\br{x}
    \brf{
        x + \frac{1}{2} + \ln\br{x} + 2\ln\br{\frac{1-x}{x}}
    }
    \nn\\
    &+&
    \int\limits_x^1
    dy \frac{\br{y^2+x^2}}{y^2\br{y-x}}
    \brs{\Phi_H\br{y}-\Phi_H\br{x}}.
    \label{IHPrecise}
\ea
The contribution of the QED corrections (\ref{HQEDCorrection})
is shown in Fig.~\ref{FigHQCDvsQED} and compared with the
QCD corrections from (\ref{HQCDRC}).
In Fig.~\ref{FigTauSpectrum} we show the lepton energy spectrum in the Born
approximation and compare it with the radiatively corrected one.
Contrasting the $\tau$-energy spectra in this figure with the corresponding
spectra in  Fig.~\ref{FigElectronSpectrum} shows that the $\tau$-leptons from the
decay $t \to b H^+ \to b \tau^+ \nu_\tau$ are distinctly more energetic. This
feature is well known in the literature. We have calculated here the (QCD + QED)
corrections to these spectra and checked their perturbative stability.
%
%
\section{Top decay channels involving the $\tau$ lepton}
\label{TauSpectrumModification}
%
The radiatively corrected charged lepton energy spectra from the decays
$t \to b W^+ \to b \ell^+\nu_\ell$ (for $\ell^+=e^+,\mu^+,\tau^+$) and
$t \to b H^+ \to b (\tau^+\nu_\tau)$ presented here will be helpful  in undertaking
 precision tests of the SM and in the searches for the $H^\pm$-induced effects in the
semileptonic decays of the top quark. Integrating these spectra from some experimental
threshold lepton energy, the anticipated enhancement in the branching ratio for the
 $t \to b (\tau^+\nu_\tau)$ mode over the other two
semileptonic modes $t \to b (\mu+\nu_\mu, e^+\nu_e)$ provides the
experimental handle on the $H^\pm$ searches. This is the strategy which is being used
at the Tevatron, where searches have also been made in the decays $H^+ \to
c\bar{s}$ (and in the charge conjugate modes), but this final state is of interest
only in the region $\tan \beta < 1$, which we do not entertain here. However, as
already mentioned in the introduction, the characteristic polarisation of the $\tau^\pm$
produced in the decays $W^\pm \to \tau^\pm \nu_\tau$ and $H^\pm \to \tau^\pm \nu_\tau$,
which reflects itself in the energy distributions of the $\tau^\pm$-decay products, can
be used to discriminate the $W^\pm$-induced and $H^\pm$-induced final states. In this
section, we calculate the energy spectra of
 the so-called single charged-prong events in $\tau$-decays.
 The $\tau^\pm$-polarisation effects  on the $\tau^\pm$ decay products
have already been investigated in the literature, in particular in
 \cite{Bullock:1992yt,Raychaudhuri:1995cc}, which we shall make use of,
 convoluting these spectra with the $\tau^+$-energy spectrum from the decay chain
 $t \to b (W^+,H^+) \to b (\tau^+ \nu_\tau)$ calculated by us here.
To that end, we  consider the following $\tau^\pm$ decay chains
\ba
t&\to& b(W^+,H^+)\tau \bar{\nu}_\tau \to jet(b) +\bar{\nu}_\tau+\nu_\tau+\bar{\nu}_l+l,
\qquad  l=e^+,\mu^+; \label{TauDecayChannels} \\
t&\to& b(W^+,H^+)\tau \bar{\nu}_\tau \to jet(b) +\bar{\nu}_\tau+\nu_\tau+\pi^+; \nn\\
t&\to& b(W^+,H^+)\tau \bar{\nu}_\tau \to jet(b) +\bar{\nu}_\tau+\nu_\tau+\pi^++\pi^0; \nn\\
t&\to& b(W^+,H^+)\tau \bar{\nu}_\tau \to jet(b) +\bar{\nu}_\tau+\nu_\tau+\pi^++2\pi^0; \nn\\
t&\to& b(W^+,H^+)\tau \bar{\nu}_\tau \to jet(b) +\bar{\nu}_\tau+\nu_\tau+2\pi^++\pi^-, \nn
\ea
involving the leptons $e^+$, $\mu^+$, the $\pi^+$, the vector and the axial-vector mesons
 $\rho^+$ and $a_1^+$,
respectively,  with the subsequent decays of the $\rho^+$ and $a_1^+$, as indicated.
Keeping in mind the long-distance nature of the QED interactions, providing the "large logarithms",
one must include the structure function associated  factors only with the final charged particles-leptons
or pions
(see Fig.~\ref{FigSFKinematicsLeptons}, b and Fig.~\ref{FigSFKinematicsPions}, a, b).
In the rest frame of the top quark, the $\tau$-leptons from the decays
$t \to b (W^+, H^+) \to b (\tau^+ \nu_\tau)$ have much larger energy and 3-momentum
compared to the$\tau$- mass, i.e. $E_\tau \gg m_\tau$.
The energy spectrum of the $\tau$-lepton decay products must be modified to take this into
account~\cite{Bullock:1992yt}. For example for
the decay $\tau\to\mu \nu\nu$ of the
$\tau$-lepton with energy $E_\tau$, the $E_\mu$-energy spectrum can be obtained from
 (\ref{BornMuonDecayWithPolarization})
(see also Eq.~(2.8) in \cite{Bullock:1992yt}):
\ba
\frac{dBr_B^{\tau\to\mu \nu\nu}}{dz}&=&\int d \cos\theta \int\limits_z^1\frac{dBr_B}{d\cos\theta d x}
\delta\br{z-\frac{x}{2}\br{1-\cos\theta}} dx \nn\\
&=& \Phi_l\br{z} = \frac{B_l}{3}\br{1-z}\brs{5+5z-4z^2-P_\tau\br{1+z-8z^2}},
\label{TauDecayLeptonSpectrum}
\ea
where $z$ is the energy fraction of the $\mu$ in the indicated decay:
\ba
z=\frac{E_a}{E_\tau}=\frac{x}{2}(1-\cos\theta) = \frac{x_a}{x_\tau},
\ea
$x=2E_\mu/m_\tau$, and $\theta$ is the angle between the directions of the $\tau$ and
the $\mu$ 3-momenta.
The index $a$ in $E_a$ shows the final particle involved. Here, $a=\mu$ (the expression above holds also
for the $e^+$-energy spectrum). We
also need the energy spectra for other particles  in the $\tau$-lepton decay, i.e. for
$a = \pi^+,\rho,a_1$. The corresponding distributions were obtained in~\cite{Bullock:1992yt}:
 for $\Phi_\pi\br{z}$ see Eq.~(2.4)
and for $\Phi_{\rho,a_1}\br{z}$ see Eq.~(2.22) in the cited paper.

%
\subsection{Leptons in final state}

\begin{figure}
    \includegraphics[width=0.8\textwidth]{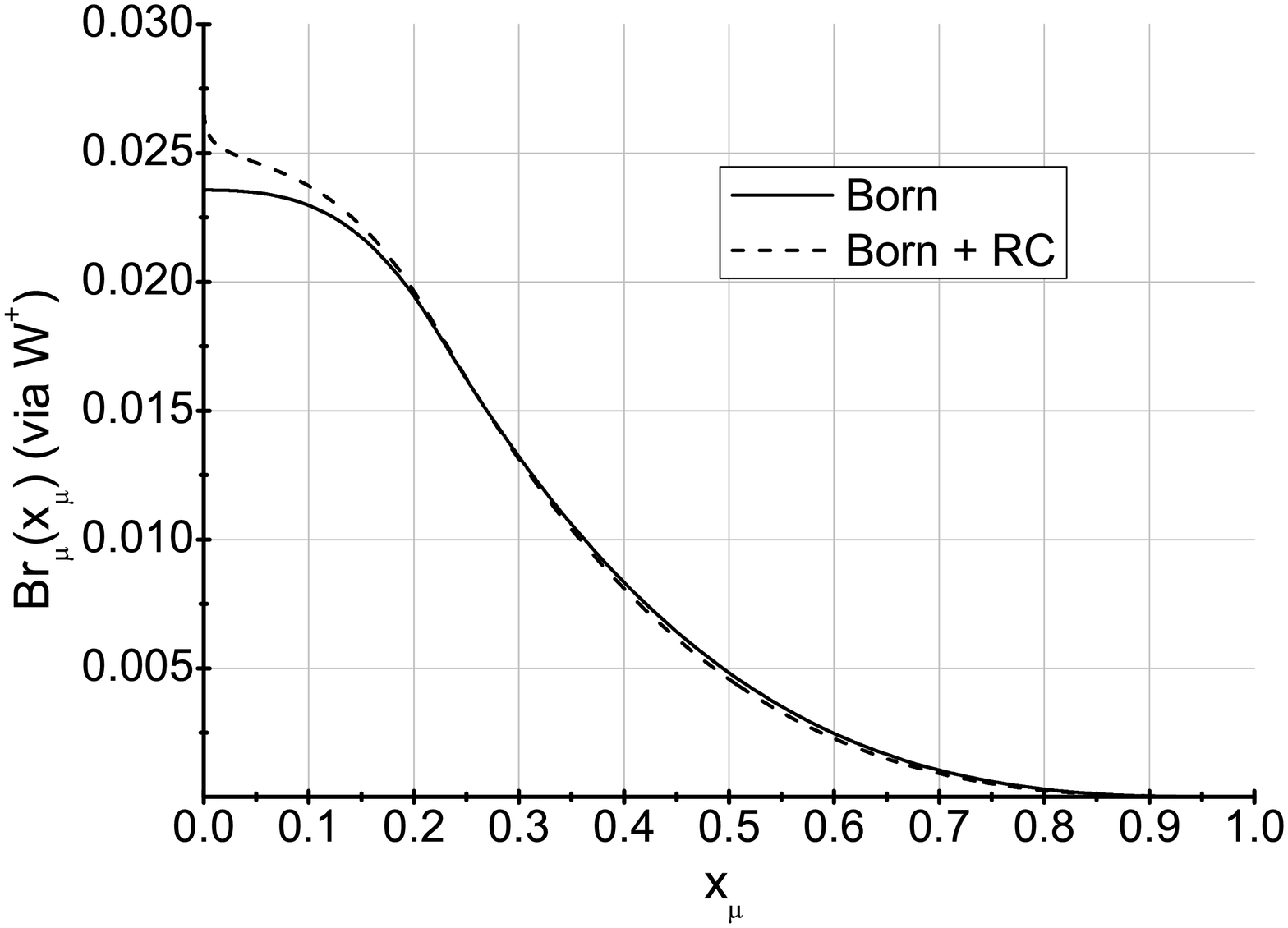}
    \caption{\label{FigLeptonW}
    Differential branching ratio
    $\frac{dBr^{t\to b W^+ \to b ((\tau^+\to\mu^+\nu_\mu \bar \nu_\tau)\nu_\tau}_B}{dx_\mu}$
    (see (\ref{BullockLepton})) as a function of the $\mu^+$ energy fraction
    $x_\mu$ in the Born approximation (solid curve) and  with the QED and QCD radiative corrections
    (dashed curve) (see (\ref{BullockLeptonWithRC})). }
\end{figure}

\begin{figure}
    \includegraphics[width=0.8\textwidth]{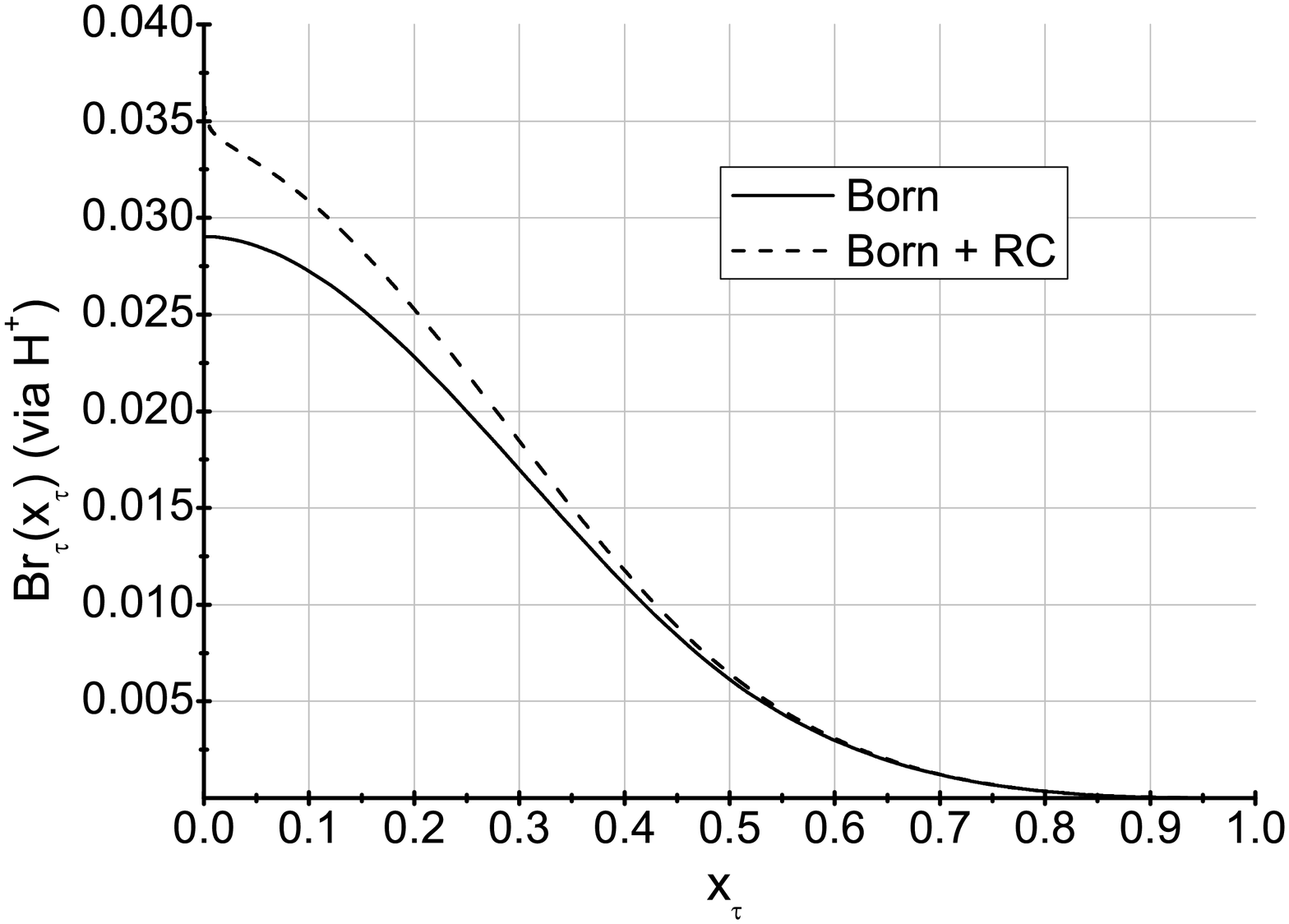}
    \caption{\label{FigLeptonH}
    Differential branching ratio
    $\frac{dBr^{t\to b H^+ \to b ((\tau^+\to\mu^+\nu_\mu \bar \nu_\tau) \nu_\tau}_B}{dx_\mu}$
     as a function of the $\mu^+$
 energy fraction $x_\mu$ in the Born approximation (solid curve)
 and with the QED and QCD radiative corrections
    (dashed curve), involving the $b H^+$  intermediate state.}
    \end{figure}

\begin{figure}
    \includegraphics[width=0.8\textwidth]{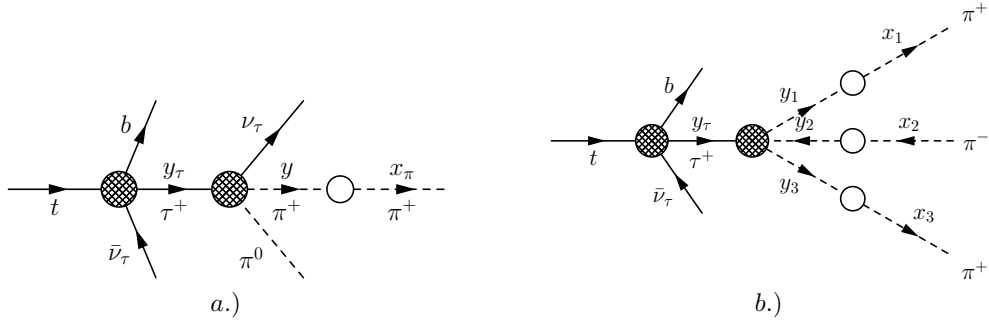}
    \caption{\label{FigSFKinematicsPions}
    Kinematics of the Structure Function method for the decay a) $ t \to b \bar{\nu}_\tau \nu_\tau \pi^0 \pi^+$
    and b) $ t \to b \bar{\nu}_\tau \nu_\tau \pi^+ \pi^- \pi^+$, involving a $\pi^+$ in the final state.}
\end{figure}

For the $\tau^+$-decay with the leptons in the final state (i.e. $a = e^+,\mu^+$),
the final expression for the lepton energy spectrum in the Born approximation is
\ba
    \frac{dBr^{t \to b +\bar{\nu}_\tau+\nu_\tau+\nu_l+\ell^+}_B}{dx_\ell}
    =
    \int\limits_{x_l}^{1} \frac{dy_\tau}{y_\tau}
    \frac{dBr^{t\to b (\tau^+ \nu_\tau)}_{\rm Born}}{dy_\tau}
    \Phi_l\br{\frac{x_l}{y_\tau}}
    \label{BullockLepton}
\ea
where the $t$-quark decay width
$\frac{dBr^{t\to b (\tau^+ \nu_\tau)}_{\rm Born}}{dx_\tau}$
is taken from (\ref{ElectronSpectrumBorn}) or (\ref{TauSpectrumBorn})
and $\Phi_l\br{z}$ was defined in (\ref{TauDecayLeptonSpectrum}).
To take into account the QED radiative corrections in LLA
we again use the SF approach illustrated in Section~\ref{MuonSection}
(see formula (\ref{MuonSpecrtumWithRC}) and Fig.~\ref{FigSFKinematicsLeptons}, b).
The radiatively corrected spectrum is then given by
\ba
    \frac{dBr^{t \to b +\bar{\nu}_\tau+\nu_\tau+\nu_\ell+\ell^+}_{QED_{LLA}+QCD}}{dx_l}
    &=&
    \frac{\brf{1,B_\tau^H}}{\Gamma_{t,RC}^{tot}}
    \int\limits_{x_l}^1 \frac{dy}{y}
    D_l\br{\frac{x_l}{y},\beta_l}
    \int\limits_{2\sqrt{\eta}}^1 dx_b
    \times\nn\\
    &\times&
    \frac{d\Gamma^{t \to b +\bar{\nu}_\tau+\nu_\tau+\nu_\ell+\ell^+}_{\rm Born}}{dx_b dy}
    \br{1-\frac{\alpha_s}{\pi}
    \brf{
            \frac{2}{3}
            \frac{F_W\br{y,x_b}}{y\br{y^{max}-y}},
            F_H\br{y,x_b}
        }
    },
    \label{BullockLeptonWithRC}
\ea
where the first entry in the curly braces is for the decay channels which go via the
$W^+ b$ intermediate state, and the second entry is  for the decays which go via
the intermediate state $H^+ b$.
The function $F_W\br{x_l,x_b}$ represents the non-leading
contributions of the QCD corrections and was defined in (\ref{Fdef}).
The definition of the function $F_H\br{x_l,x_b}$ is given in (\ref{DefFH}).
The differential branching ratios from the decay chain
 $t \to bW^+ \to b (\tau^+ \to \mu^+ \nu_\mu \bar \nu_\tau)\nu_\tau$ are shown in
Fig.~\ref{FigLeptonW} and from the decay chain
$t \to bH^+ \to b (\tau^+ \to \mu^+ \nu_\mu \bar \nu_\tau)\nu_\tau$ in Fig.~\ref{FigLeptonH}.

\subsection{Hadrons in the final state}

In this case, we have to take into account the decay chain involving the final decays
 $\tau^+ \to (\pi^+, \rho^+, a_1^+) \nu_\tau$  with the
 subsequent decays of the $\rho^+$- and the $a_1^+$-mesons into pions. The $\tau$-energy
spectrum is already given above.
Consider first the decay  $\tau^+ \to \rho^+ \bar{\nu}_\tau$, in which case the
$\rho^+$-energy spectrum is given by
\ba
    \frac{dBr^{t \to b +\nu_\tau+(\tau^+ \to \rho^+ + \bar{\nu}_\tau)}_B}{dx_\rho}
    &=&
    \int\limits_{x_\rho}^1 \frac{dx_\tau}{x_\tau}
    \frac{dBr^{t \to b +\bar{\nu}_\tau+\tau}_B}{dx_\tau}
    \Phi_\rho\br{\frac{x_\rho}{x_\tau}},
\ea
where the function $\Phi_\rho(x)$ describes the conversion of the energetic $\tau$-lepton
into the energetic $\rho$-meson. This function was calculated in Ref.~\cite{Bullock:1992yt},
which we incorporated in our numerical calculations.
The distribution in the pion energy fraction $x_\pi$ resulting from the
 $\rho^+ \to \pi^+\pi^0$ decay takes the form:
\ba
    \frac{dBr^{t \to b +\nu_\tau+(\tau \to \bar{\nu}_\tau+ (\rho\to \pi^+ \dots))}_{\rm Born}}{dx_\pi}
    &=&
    \int\limits_{x_\pi}^1 \frac{dx_\rho}{x_\rho}
    \frac{dBr^{t \to b +\nu_\tau+(\tau \to \rho+\bar{\nu}_\tau)}_{\rm Born}}{dx_\rho}
    R_\rho\br{\frac{x_\pi}{x_\rho}},
    \label{RoyBullockSpectra}
\ea
where the function $R_\rho(x)$ describes the conversion of the energetic $\rho$ meson into energetic pions
(i.e. $\rho^\pm\to\pi^\pm\pi^0$):
\ba
    R_\rho\br{x} = \frac{1}{\Gamma}
    \frac{d\Gamma}{dx},
    \qquad
    x = \frac{E_{\pi^\pm}}{E_\rho}
      = \frac{x_{\pi^\pm}}{x_\rho},
\ea
This function was investigated in \cite{Raychaudhuri:1995cc} (see Fig.~1 in \cite{Raychaudhuri:1995cc}).

The radiative corrections ("large distance contributions") can be obtained by using the
structure function approach as:
\ba
    \frac{dBr^{t \to b +\nu_\tau+(\tau \to \bar{\nu}_\tau+ (\rho\to \pi^+ \dots))}_{RC}}{dx_\pi}
    &=&
    \frac{\brf{1,B_\tau^H}}{\Gamma_{t,RC}^{tot}}
    \int\limits_{x_\pi}^1 \frac{dy_\pi}{y_\pi}\times \nn\\
    &\times&
    \frac{d\Gamma^{t \to b +\nu_\tau+(\tau \to \bar{\nu}_\tau+ (\rho\to \pi^+ \dots))}_{\rm Born}}{dy_\pi}
    D_\pi\br{\frac{x_\pi}{y_\pi},\beta_\pi},
    \label{RoyBullockSpectraWithRC}
\ea
where $D_\pi(z)$ is the structure function of the charged pion \cite{Kuraev:2006yq}:
\ba
    D_\pi\br{x,\beta_\pi} &=&
    \delta\br{1-x} + \beta P_\pi^{(1)}\br{x} +
    \frac{1}{2!} \beta^2 P_\pi^{(2)}\br{x} + \cdots,
    \\
    \beta_\pi &=& \frac{\alpha}{2\pi}\br{\ln\frac{m_t^2}{M_\pi^2}-1}, \nn
\ea
and the quantities $P_\pi^{(n)}\br{x}$ have the form:
\ba
    P_\pi^{(1)}\br{x} &=&
    \br{\frac{2x}{1-x}}_+
    =
    \lim_{\Delta\to 0}
    \brs{
        \frac{2x}{1-x} \theta\br{1-x-\Delta}
        +
        \br{2\ln\br{\Delta} + 2}\delta\br{1-x}
    },
\ea
The formula similar to (\ref{IHPrecise}) in the case of pions reads as:
\ba
    \int\limits_{x}^1 \frac{dy}{y} P_\pi^{(1)}\br{\frac{x}{y}} \phi\br{y}
    &=&
    \phi \br{x}
    \brf{
        2\ln\br{1-x} + 2
    }
    +
    \int\limits_x^1
    \frac{dy}{y} \frac{2x}{y-x}
    \brs{\phi\br{y}-\phi\br{x}},
\ea
where $\phi(x)$ is any arbitrary function we want to convolute with
the structure function $D_\pi\br{x,\beta_\pi}$ in the leading order perturbation theory.
The smoothed form of the structure function $D_\pi\br{x,\beta_\pi}$ takes the form:
\ba
    D_\pi\br{x,\beta_\pi} = 2\beta_\pi
    \br{1-z}^{2\beta_\pi-1}\br{1+\beta_\pi} - \beta_\pi
    + O\br{\beta_\pi^2}.
\ea

The  energy distribution of the charged pion obtaining from the decay
$a_1^+ \to \pi^+...$ from the parent $\tau$ decay can be derived analogously.

\begin{figure}
    \includegraphics[width=0.8\textwidth]{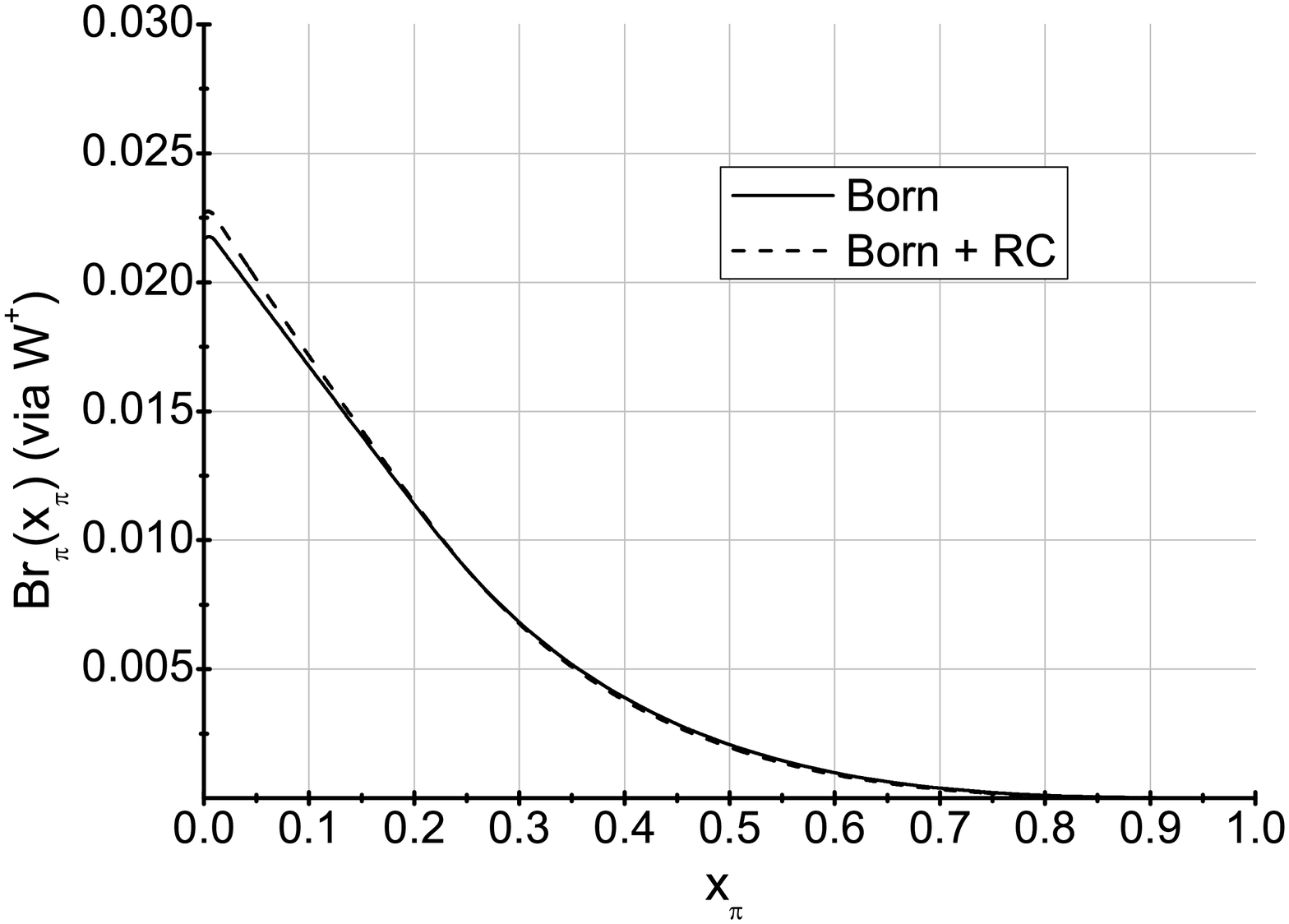}
    \caption{\label{FigPionW}
     Differential branching ratio
    $\frac{dBr^{t\to b W^+ \to b ((\tau^+ \to \pi^+ \bar \nu_\tau)\nu_\tau}_B}{dx_\pi}$ (see (\ref{BullockLepton})
    as a function of the $\pi^+$ energy fraction
    $x_\pi$ in the Born approximation (solid curve) and  with the QED and QCD radiative corrections
        taken into account (dashed curve) (see (\ref{BullockLeptonWithRC})). }
\end{figure}

\begin{figure}
    \includegraphics[width=0.8\textwidth]{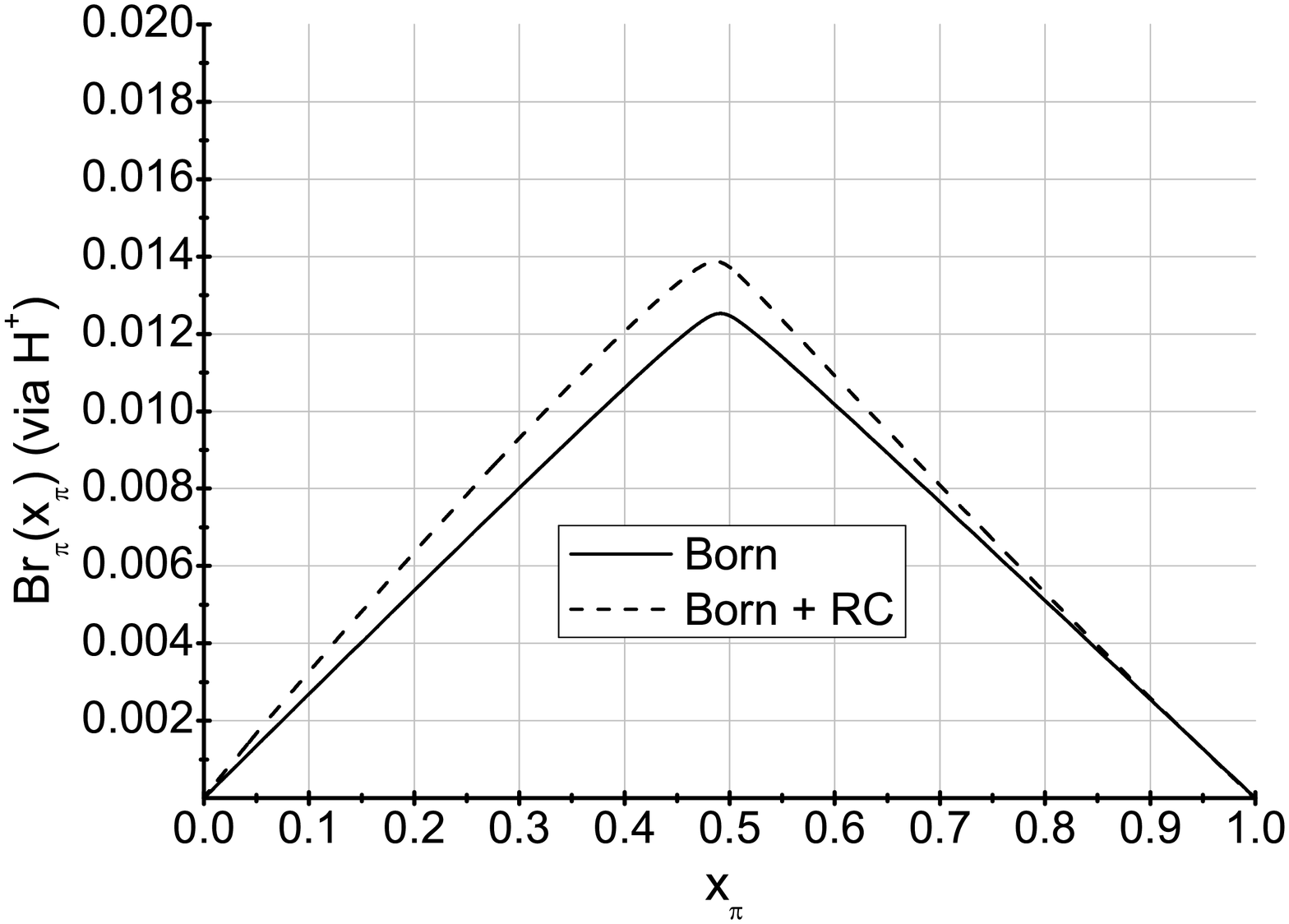}
    \caption{\label{FigPionH}
    Differential branching ratio
    $\frac{dBr^{t\to b H^+ \to b ((\tau^+ \to \pi^+ \bar \nu_\tau)\nu_\tau}_B}{dx_\pi}$
    as a function of the $\pi^+$ energy fraction
    $x_\pi$ in the Born approximation (solid curve),
    and with the QED and QCD radiative corrections
        taken into account (dashed curve), involving the $b H^+$  intermediate state.
        }
\end{figure}

\begin{figure}
    \includegraphics[width=0.8\textwidth]{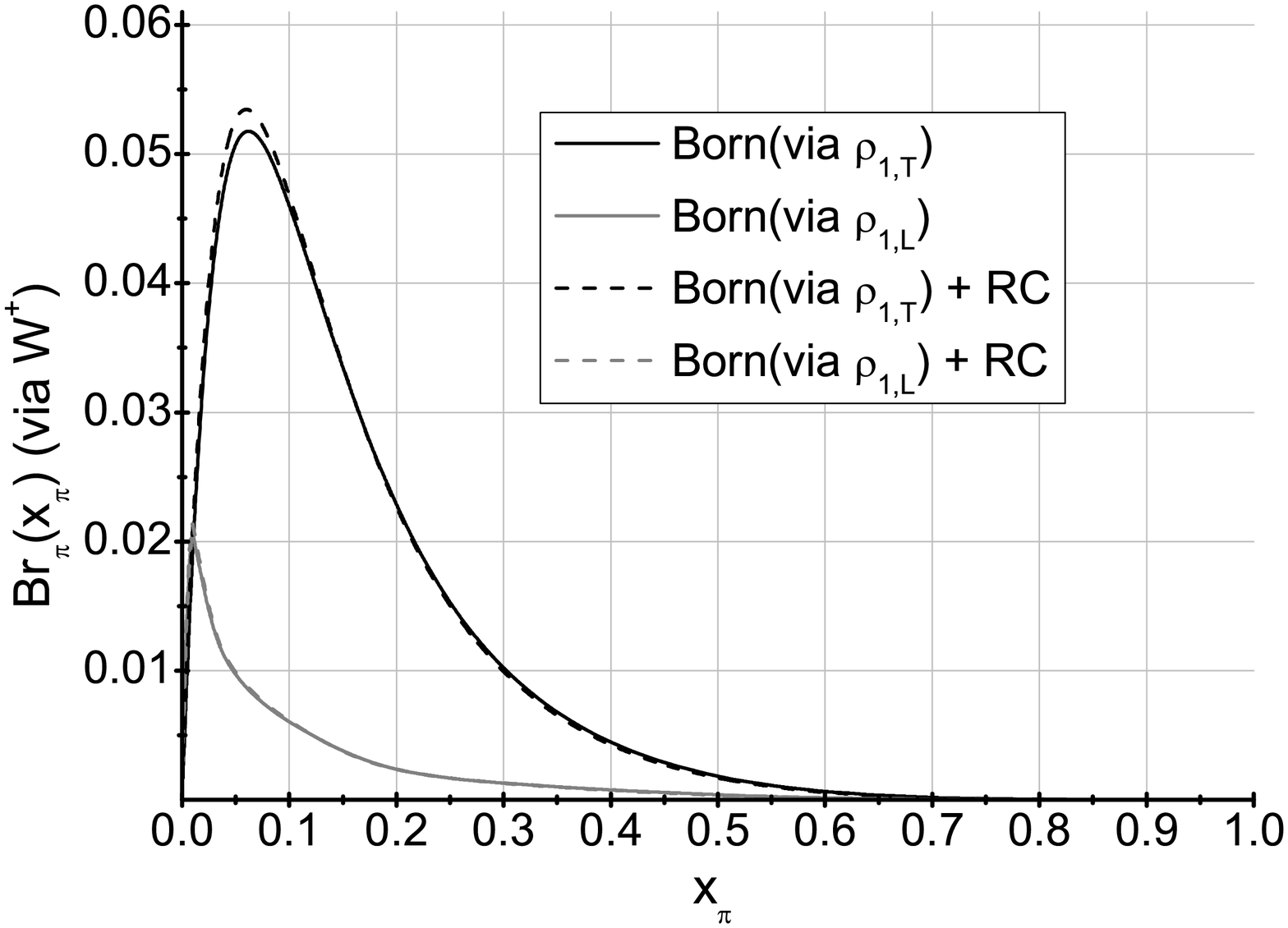}
    \caption{\label{FigRhoW}
     Differential branching ratio
       $\frac{dBr^{t\to b W^+ \to b ((\rho^+ \to \pi^+\pi^0) \bar \nu_\tau)}_B}{dx_\pi}$
     as a function of the pion energy fraction
    $x_\pi$ in the Born approximation (solid curves) and including the QED and QCD radiative
     corrections  (dashed curves). The upper curves are for the transverse polarisation of
      the $\rho$ and the lower curves correspond to the case where the
      $\rho$ is longitudinally polarised.}
\end{figure}

\begin{figure}
    \includegraphics[width=0.8\textwidth]{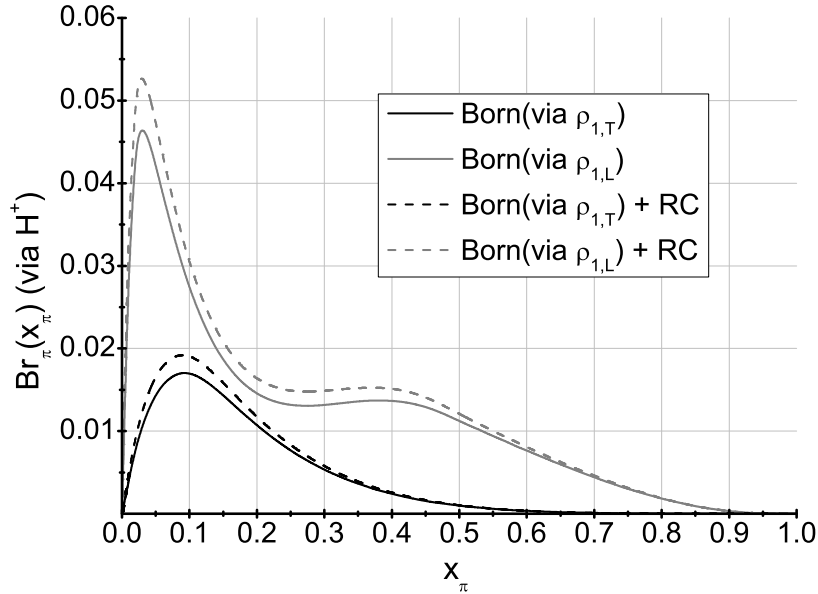}
    \caption{\label{FigRhoH}
      Differential branching ratio
       $\frac{dBr^{t\to b H^+ \to b ((\rho^+ \to \pi^+\pi^0) \bar \nu_\tau)}_B}{dx_\pi}$
      as a function of the pion energy fraction
    $x_\pi$ in the Born approximation (solid curves) and including the QED and QCD radiative
    corrections  (dashed curves), involving the $b H^+$  intermediate state. The upper curves
     are for the longitudinal polarisation of the $\rho$ and the lower curves correspond to 
     the case where the $\rho$ is transversely polarised.
    }
\end{figure}

\begin{figure}
    \includegraphics[width=0.8\textwidth]{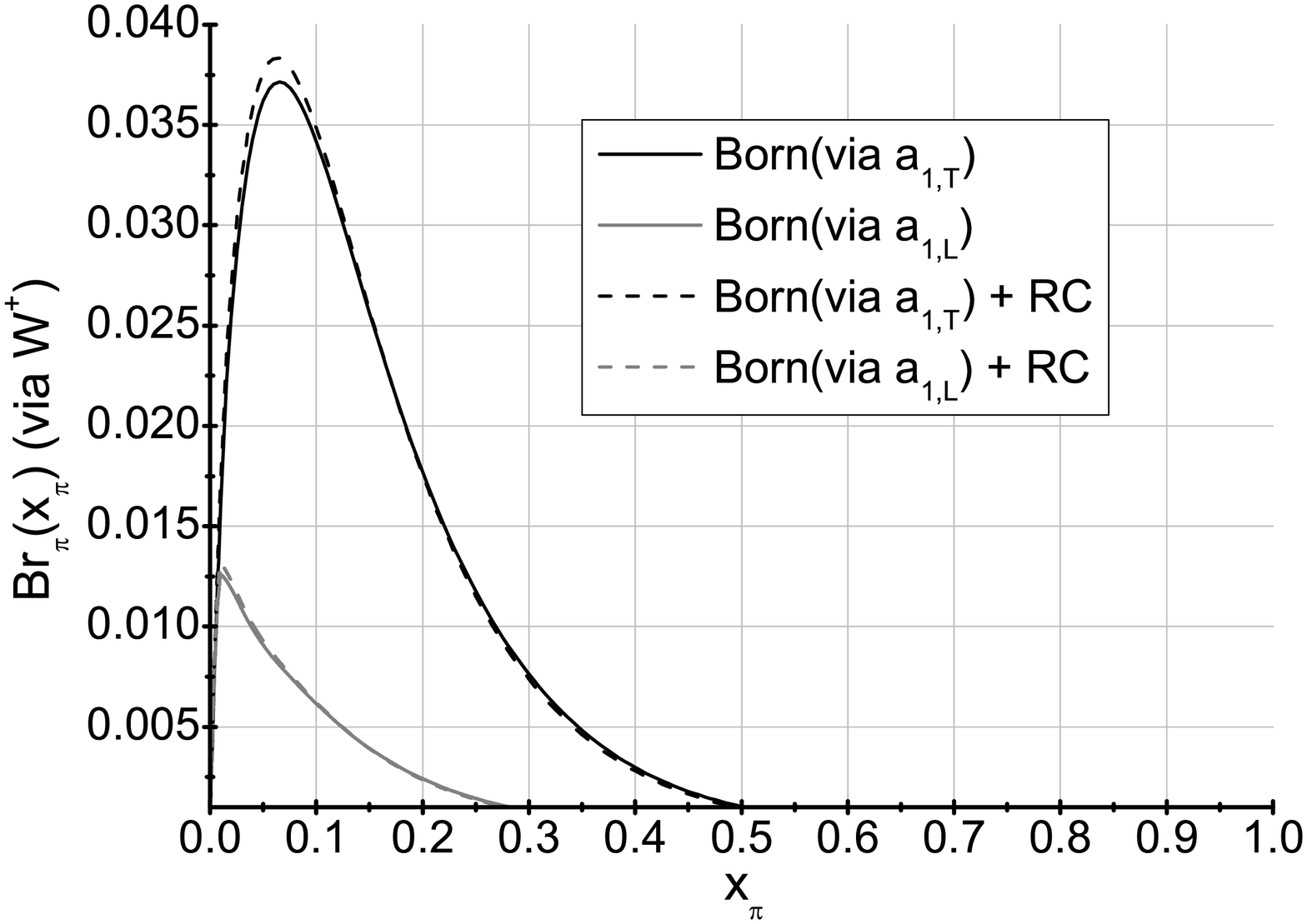}
    \caption{\label{Figa1W}
    Differential branching ratio
       $\frac{dBr^{t\to b W^+ \to b ((a_1^+ \to (3\pi)^+) \bar \nu_\tau)}_B}{dx_\pi}$
         as a function of the pion energy fraction
    $x_\pi$ in the Born approximation (solid curves) and including the QED and QCD radiative
    corrections (dashed curves). The upper curves
     are for the transverse polarisation of the $a_1$ and the lower curves correspond to
      the case where the $a_1$ is longitudinally polarised. }
\end{figure}

\begin{figure}
    \includegraphics[width=0.8\textwidth]{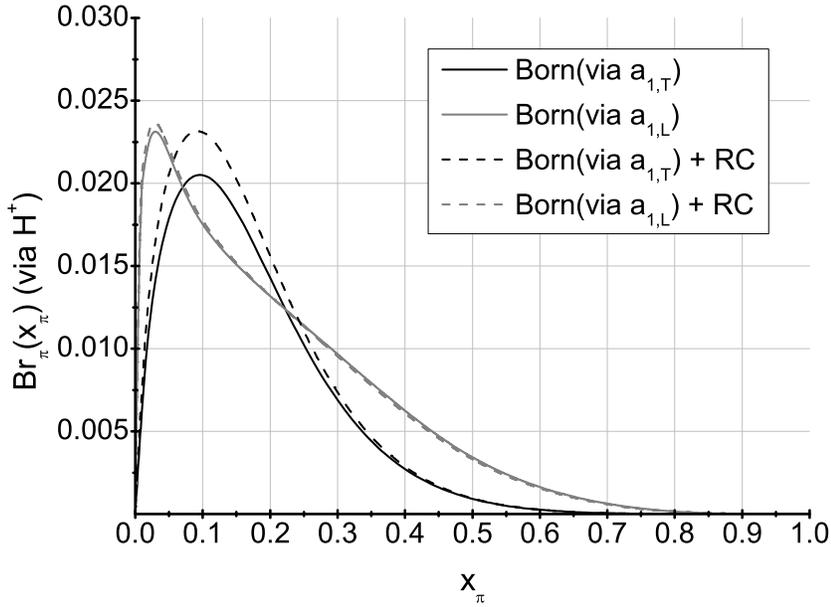}
    \caption{\label{Figa1H}
Differential branching ratio
       $\frac{dBr^{t\to b H^+ \to b ((a_1^+ \to (3\pi)^+) \bar \nu_\tau)}_B}{dx_\pi}$
         as a function of the pion energy fraction
    $x_\pi$ in the Born approximation (solid curves) and including the QED and QCD radiative corrections
    (dashed curves), involving the $b H^+$  intermediate state.
 The polarisation of the $a_1^+$ is indicated in the figure. }
\end{figure}

\begin{figure}
    \includegraphics[width=0.8\textwidth]{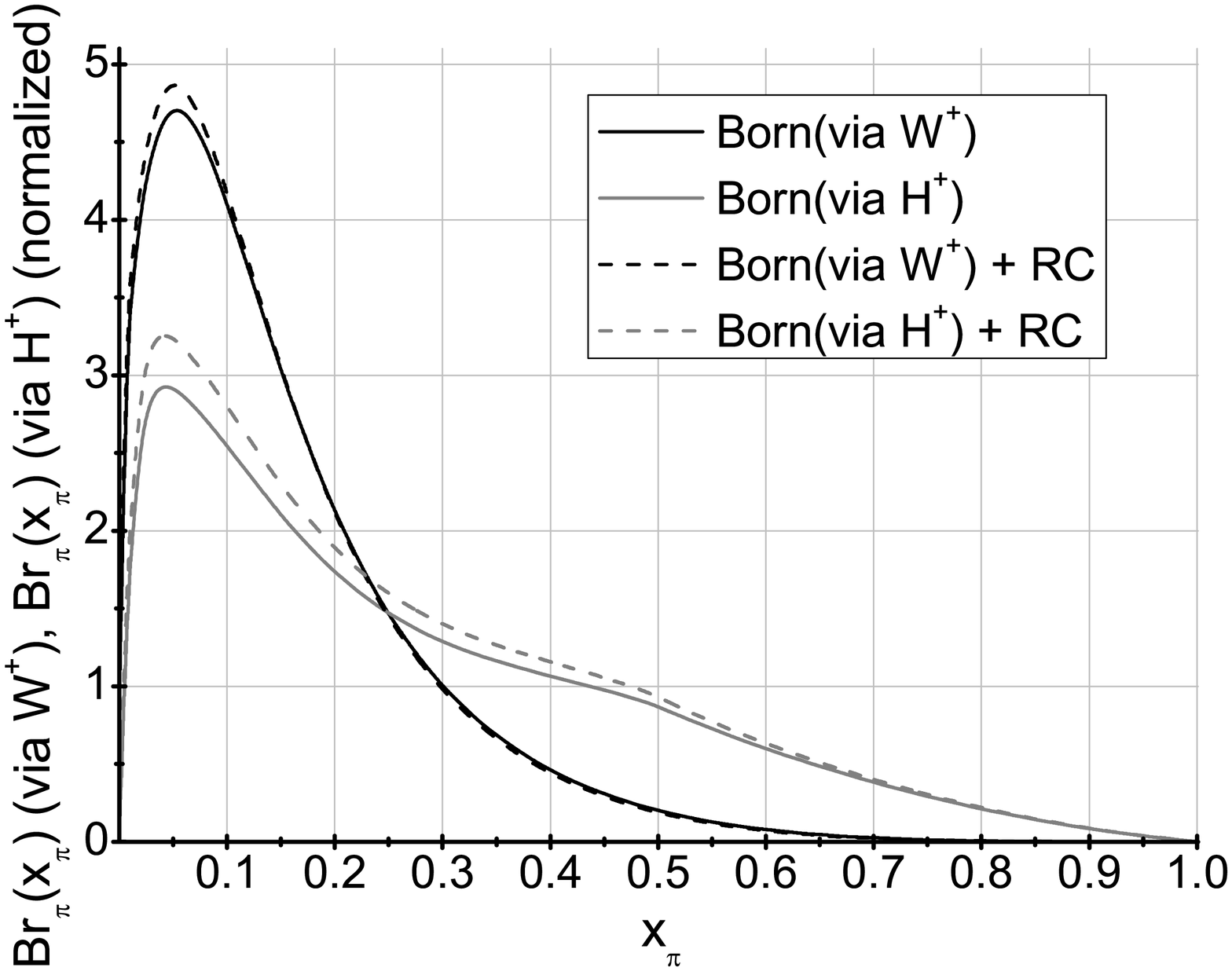}
    \caption{\label{FigNormalizedPionDistributions}
    Normalised inclusive pion energy spectra from the decays
 $t \to b (W^+,H^+) \to b (\pi^+ +X) \bar \nu_\tau$
    as a function of the pion energy fraction
    $x_\pi$ in the Born approximation (solid curves) and including the QED and QCD radiative
     corrections (dashed curves).}
\end{figure}
\section{Summary}
\label{TopHiggsSummary}

In the first part of our paper, we have calculated the QCD and QED radiative corrections
 to the semileptonic decays
$t \to b W^+ \to b \ell^+\nu_\ell$ $(\ell=e, \mu, \tau)$ in the SM. Of particular interest
are the charged lepton energy spectra, which we have calculated using
the SF approach to resum the leading order (in QED) and leading and next-to-leading order
(in QCD) contributions. These spectra will be measured accurately at the
 LHC and will be crucial to check the lepton ($e, \mu, \tau$) universality in the semileptonic
decays of the top quarks in SM. In doing this, it will be crucial to take into account the QED and
QCD radiative corrections in the energy spectra. The numerical extent of such corrections is shown
in Fig. \ref{FigRatio} for the ratios $R_{e \tau}$ and $R_{\mu \tau}$, which is one of
our principal results in this paper. The rest of our paper is addressed to the possible effects
of a charged Higgs boson $H^\pm$ with $M_{H^\pm} < m_t-m_b$ in the semileptonic decays of the
top quark. To avoid the constraints
on $M_{H^\pm}$ coming from the $B \to X_s \gamma$ decay, we assume that the Higgs sector is part of
a supersymmetric theory. Except for the SUSY radiative corrections, which can be effectively
taken into account by the supersymmetric renormalisation of the $b$-quark mass, there are
no other effects of the supersymmetric sector on the decay widths and distributions.
We have considered only the large-$\tan \beta$ parameter space of this model, in which
case the decays of the $H^\pm$ are dominated by the final state $H^\pm \to \tau^\pm \nu_\tau$.
In  Figs.~\ref{FigElectronSpectrum} and \ref{FigTauSpectrum},  we have contrasted the Born and
 radiatively corrected
$\tau$-lepton energy spectra from the decays $t \to b (W^+, H^+) \to \tau^+ \nu_\tau$ for a specific
choice of the parameters $M_{H^\pm}=120$ GeV and $\tan \beta =22$. While the Born level spectra are
well documented in the literature, effects of the radiative corrections on the spectra are, to the best
 of our knowledge, new results.

The contribution of an $H^\pm$ in $t$ ($\bar t$) decays, if allowed kinematically,  will enhance the
 decay rate for
$ t \to b \tau^+ \nu_\tau$ ($\bar t \to \bar b \tau^- \bar \nu_\tau$), which is the main
 $H^\pm$-search strategy at the Tevatron. However, with a much larger $t \bar{t}$
cross section and the luminosity anticipated at the LHC, this search strategy can be
further strengthened by taking into account the different $\tau^\pm$-polarisations in the
decays $W^\pm \to \tau^\pm \nu_\tau$ and $H^\pm \to \tau^\pm \nu_\tau$. As the polarisation information
 of the $\tau^\pm$ is transmitted to the decay products of the $\tau^\pm$, we have calculated
the energy distributions of the charged particles ($e^\pm, \mu^\pm, \pi^\pm, \rho^\pm, a_1^\pm$)
in the  single-charge-prong decays of the $\tau^\pm$, as well as the inclusive charged pion
spectra from the decay chains $t \to  b (W^\pm, H^\pm) \to b (\tau^\pm, \nu_\tau)\to b \pi^\pm +X$.
The results at the Born level are well known in the literature. We have calculated the perturbative
stability of these distributions. The entire effects of the radiative corrections presented here
can be implemented in existing Monte Carlos, such as PYTHIA and HERWIG, to provide an improved
 theoretical profile of the semileptonic decays of the top quark in the SM and can be
combined with FeynHiggs to include the SUSY-related corrections specific
to particular MSSM scenarios.

{\bf Acknowledgments}

The work presented here is partially supported by the Heisenberg-Landau Program. We thank
Gustav Kramer for reading the manuscript and helpful comments, and Sven Moch for
 communication on the top quark production cross section at the LHC.
\appendix

\section{Numerical values of the input parameters}
\label{Parameters}

For our numerical calculations we used the following values
of the parameters:

\begin{tabular}{|c|c||c|c||c|c|}
\hline
Parameter & Value & Parameter & Value & Parameter & Value \\
\hline
   $\alpha^{-1}$ & $137.035999679$ &
    $m_t$ & $171.2\GeV$ &
    $\tan\beta$ & $40$ \\
    $\alpha_s$ & $0.1176$ &
    $m_b$ & $4.20\GeV$ &
    $Br\br{\tau\to\mu\nu_\tau\bar\nu_\mu}$ & $17.36 \times 10^{-2}$ \\
    $m_e$ & $0.510998910\MeV$ &
    $M_\pi$ & $0.13957018\GeV$ &
    $Br\br{\tau\to\pi\nu_\tau}$ & $10.91 \times 10^{-2}$ \\
    $m_\mu$ & $105.6583668\MeV$ &
    $M_\rho$ & $0.775\GeV$ &
    $Br\br{\tau\to\rho\nu_\tau}$ & $25.52 \times 10^{-2}$ \\
    $m_\tau$ & $1.77684\GeV$ &
    $\Gamma_\rho$ & $0.1462\GeV$ &
    $Br\br{\tau\to a_1\nu_\tau}$ & $1.859 \times 10^{-1}$ \\
    $M_W$ & $80.398\GeV$ &
    $M_{a_1}$ & $1.230\GeV$ &
    & \\
    $\Gamma_W$ & $2.141\GeV$ &
    $\Gamma_{a_1}$ & $0.420\GeV$ &
    & \\
    $M_H$ & $120\GeV$ &
    $G_F$ & $1.16637 \times 10^{-5}\GeV^{-2}$ &
    & \\
    $\Gamma_H$ & $2\GeV$ &
    $g$ & $0.653057$ &
    & \\
\hline
\end{tabular}
\section{Radiative corrections to top quark decay via charged Higgs}
\label{RCToHiggs}

Here we give details of the QCD radiative corrections to the width of $t$-quark decay
$t(p)\to b(p_b)+\tau(p_\tau)+\bar{\nu}(p_\nu)$ with the charged Higgs
boson in the intermediate state.

The lowest order QCD corrections can be calculated in a similar way as in QED and
in the final result one must do  the replacements
\ba
\alpha \to \alpha_s C_F, \qquad C_F=\frac{N_c^2-1}{2N_c}=\frac{4}{3},
\ea
where $N_c = 3$ is the number of quark colours.

We start from the counter-terms associated with the $t$ and $b$ quarks.
Taking them into account yields
a multiplicative renormalisation factor in the expression for the differential width
\ba
d\Gamma \to d\Gamma Z_{bt},
\ea
\ba
Z_{bt}=1-\frac{\alpha}{2\pi}\brs{\ln\frac{\Lambda^2}{m_t^2}+\frac{3}{2}\ln\frac{m_t^2}{m_b^2}+
\frac{9}{2}-2\ln\frac{m_t^2}{\lambda^2}},
\ea
where $m_t$ and $m_b$ are the masses of top and bottom quark.
 The auxiliary parameters $\Lambda$ and $\lambda$ are introduced to regularise the ultraviolet (UV)
and infrared (IR) singularities, respectively. Sometimes $\lambda$ is also dubbed as
a fictitious "photon (gluon) mass". The dependence of the decay width on these
parameters will disappear from the final result. The UV-cutoff $\Lambda$ will
 be absorbed by the coupling constant renormalisation and the IR-cutoff $\lambda$ will be cancelled
by taking into account the emission of the virtual and real gluons.

Virtual corrections associated with the vertex type Feynman diagram require the calculation of the
following integral involving the loop 4-momentum
\ba
V=\int\frac{d^4 k}{i\pi^2}\frac{\gamma_\mu (\hat{p}_b-\hat{k}+m_b)(\hat{p}-\hat{k}+m_t)\gamma^\mu}
{(k^2-\lambda^2)((p_b-k)^2-m_b^2)((p-k)^2-m_t^2)}.
\ea
Using the Feynman prescription of combining the denominators and performing the loop momentum
integration (here we must impose the ultraviolet cut-off) we arrive at
\ba
V=4 \ln\frac{\Lambda^2}{m_t m_b} -
2\br{p p_b}\int\limits_0^1\frac{d x}{p_x^2}\ln\frac{p_x^2}{\lambda^2}-
2\br{m_t+m_b}\int\limits_0^1\frac{d x}{p_x^2}\br{x m_b+(1-x) m_t},
\ea
with $p_x^2=x^2 m_b^2+(1-x)^2m_t^2+2(p p_b)x(1-x)$ and the unit matrix in the Dirac space is implied.
Below we use the explicit form of the 1-fold integrals
\ba
&&\int\limits_0^1\frac{d x}{p_x^2}\brs{1; (1-x); \ln\frac{p_x^2}{m_t^2}}
=
\left[\frac{2}{m_t^2 y}\ln\br{\frac{y m_t}{m_b}}; \right.\nn \\
&&
\left.
\frac{1}{m_t^2(1-y)}\ln y;~~
\frac{1}{m_t^2 y}
\br{\ln^2 y-\frac{1}{2}\ln^2\frac{m_t^2}{m_b^2}-2\Li{2}\br{1-\frac{1}{y}}}
\right].
\ea
The next step consists of the calculation of the contribution arising from emission of real gluons
- soft and hard ones. Standard calculation for the case of soft gluon emission
$\omega < \Delta E_b \ll m_t$
(we work in the rest frame of top quark) leads to
\ba
\frac{d\Gamma_{soft}}{d\Gamma_B}&=&
-\frac{\alpha}{4\pi^2}\int\frac{d^3 k}{\omega}
\br{\frac{p}{p k}-\frac{p_b}{p_b k}}^2\nn\\
&=&
\frac{\alpha}{\pi}
\brs{2(l-1)\ln\frac{2\Delta E}{\lambda}+1+l-l^2-\frac{\pi^2}{6}},
\qquad
l=\ln\frac{y m_t}{m_b}.
\label{QCDHSoft}
\ea
Extracting the factor $Z=1+\frac{3\alpha}{2\pi}\ln\frac{\Lambda^2}{m_t-t^2}$ and using the
ultraviolet-regularised quantities we can write
\ba
Z d\Gamma_{unren}=d\Gamma.
\ea
Collecting the contributions of the Born level and the virtual and soft real corrections, we obtain:
\ba
d\Gamma &=& d\Gamma_B[1+\frac{\alpha}{2\pi}(L-1)(2\ln\Delta+\frac{3}{2})+
\frac{\alpha}{\pi}[-\ln\Delta-\frac{5}{2}+
\frac{1}{2}\ln^2 y-\frac{5}{2}\ln y+Li_2(1-\frac{1}{y})-\nn\\
&-&\frac{\pi^2}{6}-\frac{1}{1-y}\ln y]]
\label{VirtAndSoft}
\ea
with $\Delta=\Delta E_b/E_b$, $L=\ln\frac{m_t^2 y^2}{m_b^2 \Delta}$.

Note that the term containing $\ln\Delta$ are connected with the emission from the "light"
$b$-quark and the heavy $top$-quark.

Let us now consider the emission of the hard gluon with
momenta $k=(\omega,\vec k), k_0>\Delta E$. The relevant phase volume
\ba
d\Phi_4=\frac{(2\pi)^4}{(2\pi)^{12}}\frac{d^3p_b}{2 E_b}\frac{d^3p_\tau}{2 E_\tau}
\frac{d^3p_\nu}{2E_\nu}\frac{d^3 k}{2\omega}
\delta^4(p-p_b-p_\nu-p_\tau-k), \nn
\ea
can be transformed as
\ba
d\Phi_4=\frac{m_t^4}{2^{13}\pi^6}d y x_\tau d x_\tau d x d z dO_\tau
\delta\br{1-x_\tau-x-y+z+ \frac{2p_\tau(k+p_b)}{m_t^2}},
\ea
where
\ba
    x = \frac{2\omega}{m_t}, \qquad z = \frac{2\br{k p_b}}{m_t^2},
\ea
and
\ba
    d O_\tau=\frac{2d c_1 d c_2}{\sqrt{1-c_1^2-c_2^2-c^2+2c_1c_2c}},
\ea
is the angular phase volume of the $\tau$ lepton, and
\ba
    c = \cos(\vec{k},\vec{p}_b),
    \qquad
    c_1 = \cos(\vec{k},\vec{p}_\tau),
    \qquad
    c_2 = \cos(\vec{p}_\tau,\vec{p}_b).
\ea
Explicit calculation yields
\ba
\int dO_\tau \delta\br{1-x_\tau-x-y+z+\frac{2p_\tau(k+p_b)}{m_t^2}}=\frac{4\pi}{x_\tau R}~,
\quad
R = \sqrt{(x+y)^2-4z}.
\ea
and the variable $z$ is bounded by
\ba
z_m < z < x y, \qquad z_m=\frac{x m_b^2}{y m_t^2} \ll 1.
\ea
Summed over the final spin states, the matrix element squared leads to
\ba
    \frac{d\Gamma_{uncoll}}{dx_\tau dy}
    =
    \frac{\alpha}{2\pi} \int dx x
    \int\limits_\sigma^{xy} dz
    \frac{d\Gamma\br{x+y-z}}{dx_\tau dy}\frac{1}{\br{x+y-z} R}
    \br{F_1+F_2+F_3}~,
\ea
where
\ba
    F_1 &=& \frac{y^2+(y+x)^2}{x z}-\frac{2m_b^2}{m_t^2}\frac{x+y}{z^2},\nn\\
    F_2 &=& -\frac{2y}{x^2}, \nn\\
    F_3 &=& -\frac{2}{x}\br{1+y+x}+\frac{z}{x^2}\br{2+x}.\nn
\ea

It is convenient to introduce the small auxiliary parameter $\sigma$
($z_m \ll \sigma \ll x y \sim 1$) and
extract the contribution of the collinear kinematics $\vec{k}||\vec{p}_b$.

Only $F_1$ gives the contribution in the collinear region ($z < \sigma$):
\ba
d\Gamma_{coll}=\frac{\alpha}{2\pi}\int\limits_{y(1+\Delta)}^1\frac{d t}{t}
d\Gamma\br{t}
\brs{\frac{1+\frac{y^2}{t^2}}{1-\frac{y}{t}}(L_\sigma-1)+1-\frac{y}{t}}, \label{QCDHF1LL}
\ea
with
\ba
L_\sigma=\ln\frac{m_t^2 y \sigma}{x m_b^2},
\qquad
\Delta=\frac{2\Delta E}{m_t y}. \nn
\ea
Contribution of $F_1$ from the non-collinear region can be put in the form:
\ba
\frac{\alpha}{2\pi}\int\limits_{y(1+\Delta)}^1\frac{d t}{t}
d\Gamma\br{t}
\brs{\frac{1+\frac{y^2}{t^2}}{1-\frac{y}{t}}[\ln\frac{xy}{\sigma}+
\int\limits_0^{y(t-y)}\frac{d z}{z}\Sigma(t,z)]},
\ea
with
\ba
\Sigma(t,z)=\frac{d\Gamma(t-z)t^2}{(t-z)d\Gamma(t)\sqrt{t^2-4z}}-1.
\ea
Note that the second term containing $\Sigma(t,z)$ is finite in the
limit $m_b \to 0$.

The contribution of the second term ($F_2$)  can be cast in the form:
\ba
-\frac{\alpha}{\pi}\int\limits_{y(1+\Delta)}^1\frac{d t}{t-y}\frac{y^2d\Gamma(t)}{t^2}-
\frac{\alpha}{\pi}\int\limits_y^1\frac{d t d\Gamma(t)}{(t-y)^2}\int\limits_0^{y(t-y)}d z~\Sigma(t,z).
\ea
The first term above combined with the term $-\frac{\alpha}{\pi}d\Gamma(y) \ln \Delta$
(see (\ref{VirtAndSoft})) gives a quantity which is finite in the limit $\Delta\to 0$.
 Emission from the light quark has a form predicted by the Structure Function approach.
 Combining all the contributions, we arrive at the following
  expression for the QCD corrected double (Dalitz) distribution in the variables $x_\tau$, $y$:
 \ba
 \frac{d\Gamma\br{y,x_\tau}}{dx_\tau dy}=\int\limits_y^1\frac{d t}{t}D\br{\frac{y}{t},
\tilde{\beta}(y)}
 \frac{d\Gamma\br{t,x_\tau}}{d t dx_\tau}\brf{1+\frac{\alpha_s}{\pi}F_H},
\ea
where $F_H$ is the $K$-factor which contains all the non-enhanced terms.
On integrating the $b$-quark energy fraction, the mass singularities
($\tilde{\beta}(y)=\frac{\alpha_s}{2\pi}(\ln(y^2m_t^2/m_b^2)-1)$) for $m_b\to 0$
   will disappears due to the relation
   \ba
\int\limits_0^1dy\int\limits_y^1\frac{d t}{t}D\br{\frac{y}{t},\beta}F(t)=
\int\limits_0^1F(t) d t.
   \ea
Thus, in an experimental setup involving an  averaging over the
$b$-jet production, the resulting $\tau$-meson energy spectrum fraction is described by the
following expression:
   \ba
    \frac{d\Gamma}{dx_\tau}&=&
    \int\limits_{1-x_\tau}^1 d x_b
    \frac{d\Gamma}{d x_b d x_\tau},
    \label{HQCDRC}\\
    \frac{d\Gamma}{d t d x_\tau}
    &=&
    \frac{d\Gamma_B}{d t d x_\tau}
    -
    \frac{\alpha_s C_F}{\pi}
    \int\limits_{t\br{1+\Delta}}^1
    dy
    \frac{d\Gamma_B}{d y d x_\tau}
    \frac{y^2}{t^2}\frac{1}{t-y} +\nn\\
    &+&
    \frac{\alpha_s C_F}{\pi}
    \frac{d\Gamma_B}{d t d x_\tau}
    \brs{
        - \frac{5}{2} + \frac{1}{2}\ln^2 t -
        \frac{5}{2}\ln t -\frac{\ln t}{1-t} - \zeta_2 + \Li{2}\br{1-\frac{1}{t}} -\frac{t}{2}-
        \ln\Delta
    } \nn\\
    &=&
    \frac{d\Gamma_B}{d t d x_\tau}
    \br{1-\frac{\alpha_s}{\pi} F_H\br{x_\tau,t}}. \nn
   \ea
As anticipated, this expression  does not depend on the small auxiliary parameter $\Delta\ll 1$.
Thus function $F_H\br{x_\tau, x_b}$ is defined as
   \ba
    F_H\br{x_\tau,t}
    &=&
    C_F
    \left\{
        \frac{5}{2} - \frac{1}{2}\ln^2 t +
        \frac{5}{2}\ln t + \frac{\ln t}{1-t} + \zeta_2 - \Li{2}\br{1-\frac{1}{t}} 
        +\frac{t}{2}+ \ln\Delta
    \right. \nn\\
    &+&
    \left.
    \br{\frac{d\Gamma_B}{d t d x_\tau}}^{-1}
    \times
    \int\limits_{t\br{1+\Delta}}^1
    dy
    \frac{d\Gamma_B}{d y d x_\tau}
    \frac{y^2}{t^2}\frac{1}{t-y}
    \right\}.
    \label{DefFH}
   \ea
%


\end{document}